\documentclass[%
 reprint,
superscriptaddress,
 showkeys,
 amsmath,amssymb,
 aps,
]{revtex4-2}

\usepackage{graphicx}% Include figure files
\usepackage{dcolumn}% Align table columns on decimal point
\usepackage{bm}% bold math
\usepackage{siunitx}  %SI unit
\usepackage{comment}
\usepackage{multirow}
\usepackage{array}
\usepackage{textpos}

\usepackage{float}
\usepackage{esvect}
\begin{document}

\preprint{APS/123-QED}

\title{Diamagnetic composites for high-Q levitating resonators}
%High-Q levitating composite resonators
%Or: High-Q levitating resonators from diamagnetic microparticle composites
%The composite is not high-Q, the resonator is.
% Force line breaks with \\

\author{Xianfeng Chen}
 \affiliation{Department of Precision and Microsystems Engineering, Delft University of Technology, Mekelweg 2, 2628 CD, Delft, The Netherlands}%Lines break automatically or can be forced with \\
\author{Satya K. Ammu}%
\affiliation{
Shaping Matter Lab,
Faculty of Aerospace Engineering,
Delft University of Technology,
2629 HS, Delft, The Netherlands
}%

\author{Kunal Masania}
\affiliation{
Shaping Matter Lab,
Faculty of Aerospace Engineering,
Delft University of Technology,
2629 HS, Delft, The Netherlands
}%
\author{Peter G. Steeneken}
\affiliation{Department of Precision and Microsystems Engineering, Delft University of Technology, Mekelweg 2, 2628 CD, Delft, The Netherlands}
\affiliation{Kavli Institute of Nanoscience, Delft University of Technology, Lorentzweg 1, 2628 CJ, Delft, The Netherlands.}

\author{Farbod Alijani}
%\email{f.alijani@tudelft.nl}
\affiliation{Department of Precision and Microsystems Engineering, Delft University of Technology, Mekelweg 2, 2628 CD, Delft, The Netherlands}

\date{\today}% It is always \today, today,

\begin{abstract}
Levitation offers extreme isolation of mechanical systems from their environment, while enabling unconstrained high-precision translation and rotation of objects. Diamagnetic levitation is one of the most attractive levitation schemes, because it allows stable levitation at room temperature without the need for a continuous power supply. However, dissipation by eddy currents in conventional diamagnetic materials significantly limits the application potential of diamagnetically levitating systems.
Here, we present a route towards high $Q$ macroscopic levitating resonators by substantially reducing eddy current damping using graphite particle based diamagnetic composites. We demonstrate resonators that feature quality factors $Q$ above 450,000 and vibration lifetimes beyond one hour, while levitating above permanent magnets in high vacuum at room temperature. The composite resonators have a $Q$ that is more than 400 times higher than that of diamagnetic graphite plates. By tuning the composite particle size and density, we investigate the dissipation reduction mechanism and enhance the $Q$ of the levitating resonators. Since their estimated acceleration noise is as low as some of the best superconducting levitating accelerometers at cryogenic temperatures, the high $Q$ and large mass of the presented composite resonators positions them as one of the most promising technologies for next generation ultra-sensitive room temperature accelerometers.   

\end{abstract}

\keywords{Quality factor|Diamagnetic levitation|Eddy current damping|Composites}%Use showkeys class option if keyword
\maketitle

%\tableofcontents
%\vspace*{-0.2cm}
% \onecolumngrid
% %\vspace{0.2cm}
% \twocolumngrid
\section{\label{sec:intro}Introduction}
The low dissipation and high quality factor ($Q$) of mechanical resonators makes them the devices of choice in precision time-keeping, frequency filtering, and sensing applications. With the emergence of nano- and micro-electromechanical systems, and the drive towards quantum limited mechanical elements, pushing the performance boundaries of resonators has become a matter of high scientific and societal relevance \cite{maccabe2020nano,beccari2103hierarchical,kemp2019high,kumar2022activ,catano2020high,shin2022spiderweb}. In particular, mechanical energy loss via the clamping points has become a dominant factor, limiting the $Q$ of these resonators. As a consequence, attention has moved towards the field of levitodynamics \cite{brandt1989levitation,gonzalez2021levitodynamics}. By employing levitating resonators that are %Amongst different methods that can be used to boost the $Q$ of mechanical resonators \cite{miller2018effective}, high vacuum levitation through absence of clamping has gained attention recently \cite{gonzalez2021levitodynamics}. This method enables 
well isolated from their environment, losses can be minimised and extreme sensitivities can be achieved.
\vspace{0.2cm}

Optically, superconducting, and electrically levitating micro and nanoresonators have been shown to feature high $Q$s in the range $10^6-10^7$ \cite{gieseler2012subkelvin,vovrosh2017parametric,vinante2020ultralow,pontin2020ultranarrow}. %However, further Q-enhancement with these levitating schemes is hindered by several factors such as excessive heating \cite{jain2016direct, millen2014nanoscale}, drift from the applied force field \cite{bullier2020characterisation} and the need for feedback control to enable long-term stability \cite{monteiro2017optical}.
Although these techniques are of great interest for fundamental studies, the requirement for continuous position control and cooling power supply \cite{gonzalez2021levitodynamics}, narrows their application range, since the levitating object will collapse in a situation of power loss. 
Diamagnetic levitation is the only known method for realizing stable continuous vacuum levitation of objects at room temperature without external power supply \cite{simon2000diamagnetic,1335771,chen2020rigid,chen2021diamagnetically}. Moreover, unlike optical and electrical levitation that are limited to nano-gram objects \cite{monteiro2017optical,moore2021searching}, diamagnetic levitation is the method of choice for levitating macroscopic objects whose larger mass can significantly enhance the sensitivity of sensors like accelerometers  \cite{timberlake2019acceleration} and gravimeters \cite{middlemiss2016measurement,schmole2016micromechanical,marletto2017gravitationally,belenchia2018quantum}.
However, the $Q$ of conventional diamagnetic materials such as graphite that has high magnetic susceptibilities is significantly limited by eddy current damping forces \cite{chen2020rigid}. While the diamagnetic levitation of non-conductive materials such as silica could make the levitodynamic system immune to the presence of eddy current damping forces, their magnetic susceptibility is lower, such that it normally only is suitable for levitating microscopic objects \cite{slezak2018cooling,lewandowski2021high}.
\vspace{0.4cm}

%\vspace*{-0.6cm}
Here, we demonstrate millimeter scale composite plates comprising graphite microparticles dispersed in epoxy resin that levitate stably above permanent magnets and exhibit $Q$s above 450,000. The strong diamagnetic susceptibility of the graphite particles allows passive levitation of the composite plates, while the epoxy acts as an insulating material that suppresses eddy currents. To investigate the  dependence of $Q$ on composite properties, we perform simulations and experiments on composites with different particle sizes and volume fractions. We confirm that by reducing particle size, damping can be significantly decreased while maintaining the macroscopic size of the levitating object. Finally, we compare the performance of the realized diamagnetic composite resonator to state-of-the-art accelerometers and show that it leads to one of the lowest acceleration noise figures achieved thus far in levitating sensors.
\begin{figure*}
\includegraphics[height=7cm]{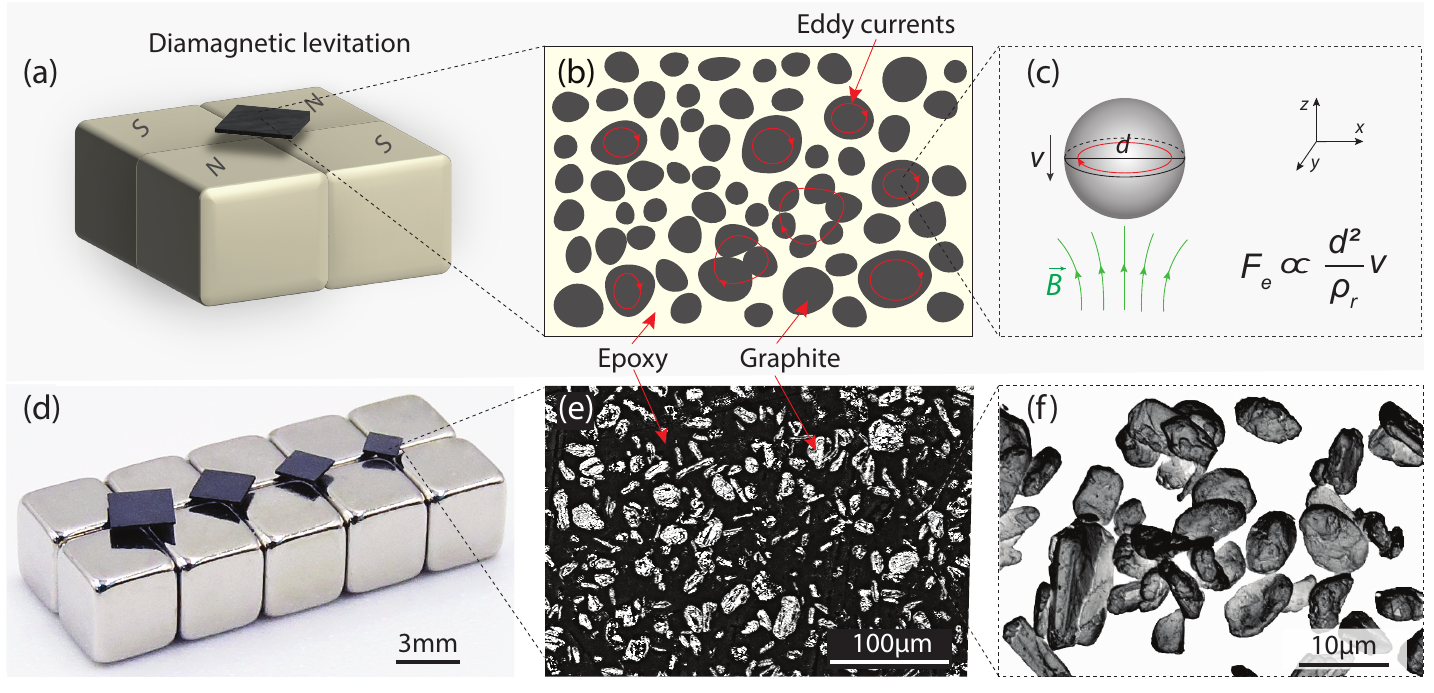}
\caption{\label{fig:concept} (a) Schematic of a diamagnetic plate levitating above 4 cubic Nd$_2$Fe$_{14}$B magnets with alternating  magnetization. (b) Schematic of the eddy currents (red  circular arrows) generated inside the graphite microparticles that are distributed in the composite. (c) The relationship between the eddy current damping force $F_\mathrm{e}$ and particle size $d$, for a spherical particle with electrical resistivity $\rho_\mathrm{r}$ moving in a magnetic field (see S5 for details). (d) An array of  graphite-epoxy composite plates of different sizes levitating above magnets at room temperature and pressure. (e) Confocal microscopy image of the surface of the composite plate with particle size of \SI{17.6}{um} and volume fraction of 0.21, showing the distribution of the graphite particles (white) in the epoxy (black). (f) Scanning electron microscopy image showing the size and morphology of the graphite particles.}
\end{figure*}

%, making it of interest for application as accelerometer\cite{timberlake2019acceleration}, gravimeter\cite{middlemiss2016measurement,schmole2016micromechanical,marletto2017gravitationally,belenchia2018quantum}, or for exploring macroscopic limits of quantum mechanics \cite{whittle2021approaching,michimura2020quantum,yu2020quantum,catano2020high}. 
%and dark matter \cite{graham2016dark,pierce2018searching,carney2021ultralight,gonzalez2021levitodynamics}. 

%for the show  The high $Q$ and mass of the composite resonator makes them strong candidates for next also suggests reduction of the acceleration noise floor down to $\SI{0.2}{ng/\sqrt{Hz}}$ at room temperature, comparable to the state-of-the-art superconducting accelerometers that operate at cryo-temperatures.  

\newpage
\section{\label{sec:results}Results}
\subsection{Diamagnetically levitating composites}
%\section{\label{sec:fab and setup} Levitating composites}

To realize diamagnetically levitating resonators with high $Q$s, we fabricate composite materials with distributed graphite microparticles by dispersing them in epoxy resin through mechanical mixing, and then curing the resin in an oven (see Methods and S1). The fabrication process enables a high degree of freedom in size of graphite particles and selection of resin composition. Due to the strong diamagnetic susceptibility of graphite, the composite levitates stably above permanent Nd$_2$Fe$_{14}$B magnets arranged in a checkerboard configuration with alternating magnetization (see Fig. \ref{fig:concept}a). We expect that the epoxy between the microparticles acts as an insulator, confining eddy currents within the particles (Fig. \ref{fig:concept}b), and thus diminishing eddy current damping forces and increasing $Q$ \cite{chen2020rigid}. Furthermore, since for a composite with particle size $d$ moving in a magnetic field, the eddy current damping force per volume scales quadratically with particle size ($F_\mathrm{e} \propto d^2$ see Fig. \ref{fig:concept}c and S5 \cite{taghvaei2009eddy}), we expect that by reducing the microparticle size in the composite, high mechanical $Q$s can be achieved while maintaining macroscopic mass. To experimentally investigate this effect, square graphite/epoxy composite plates of different size with a constant \SI{90}{\um} thickness are prepared, as shown in Fig. \ref{fig:concept}d. The successful levitation of the composite plates with graphite volume fraction $V_\mathrm{f}$ of \SI{21}{\percent}, as shown in Fig. \ref{fig:concept}d, confirms that the diamagnetism of graphite is maintained in the microparticles and that the diamagnetic force remains strong enough to oppose the gravitational force, even though the graphite particles have anisotropic magnetic susceptibilities and are randomly oriented inside the epoxy matrix. In Fig. \ref{fig:concept}e,f we show microscopic images of the composite and graphite microparticles from which we note that the particle sizes are distributed over a wide range (see the particle size measurement in S2.1). Moreover, we quantitatively analyze the particle distribution (see S2.2) and observe that the graphite particles are randomly distributed inside the epoxy matrix.  

%Moreover, to estimate the particle size we use a particle size analyzer to measure the particle size distribution of the graphite powder (see SI3).  
%\newpage
\subsection{Q-factor measurement}
\begin{figure*}
    \centering
    \includegraphics[height=5cm]{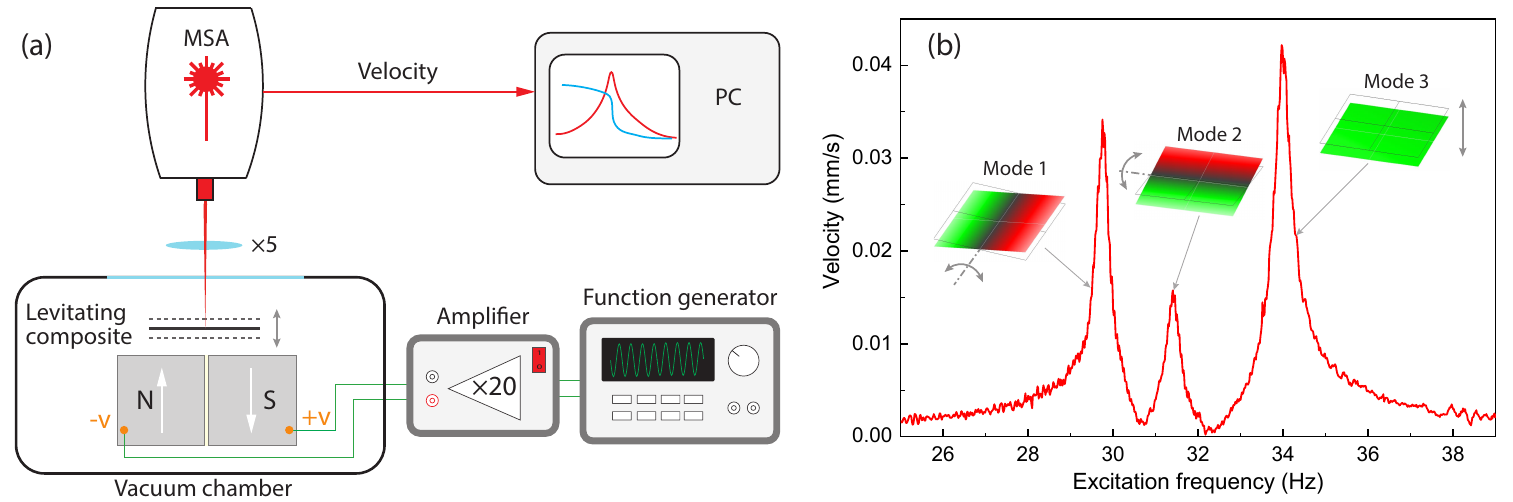}
    \caption{\label{fig:measurement-setup-and-3peaks}Experimental setup and rigid body dynamic response of a levitating composite resonator. (a) Schematic of the measurement setup comprising a MSA400 Polytec Laser Doppler Vibrometer (LDV) for the readout and electrostatic force as the actuation means. The drive voltage is generated by the function generator and is amplified by a 20$\times$ voltage amplifier that drives the levitating plate into resonance. The electrostatic force is generated by applying voltage between the magnets beneath the levitating plate. By focusing the vibrometer's laser beam on the plate, the plate motion is captured, and the acquired velocity is used for spectral analysis. (b) The frequency response curve of a $1.8\times1.8\times\SI{0.09}{mm^3}$ levitating composite plate with \SI{8.6}{\um} graphite particles measured at \SI{0.1}{mbar}. Three of the measured mode shapes using LDV are shown close to the corresponding resonance peaks.}
\end{figure*}
To probe the vibrations of the levitating plates, we use a Polytec MSA400 Laser Doppler Vibrometer (LDV) and measure their out-of-plane velocity in a vacuum chamber at a pressure of \SI{0.1}{mbar} (see Fig. \ref{fig:measurement-setup-and-3peaks}a and the Methods). We characterize the spectral response of the levitating objects by driving them electrostatically at different frequencies. Fig. \ref{fig:measurement-setup-and-3peaks}b shows the area-averaged magnitude of the spectral response for a $1.8\times1.8\times\SI{0.09}{mm^3}$ composite plate with \SI{8.6}{\um} graphite particles. Three plate resonance peaks can be identified in the spectral response, which correspond to the two rotational modes at \SI{29.7}{Hz} (Mode 1) and \SI{31.4}{Hz} (Mode 2) and the translational rigid body mode of vibration at \SI{34.0}{Hz} (Mode 3). In this work, we focus on the $Q$ of the out-of-plane translational mode that relates to the vertical motion  (Mode 3). The mode shapes are identified by scanning the laser over the plate surface at the corresponding resonance frequencies, and are shown in Fig. \ref{fig:measurement-setup-and-3peaks}b.

\begin{figure*}
    \centering
	\includegraphics[height=5cm]{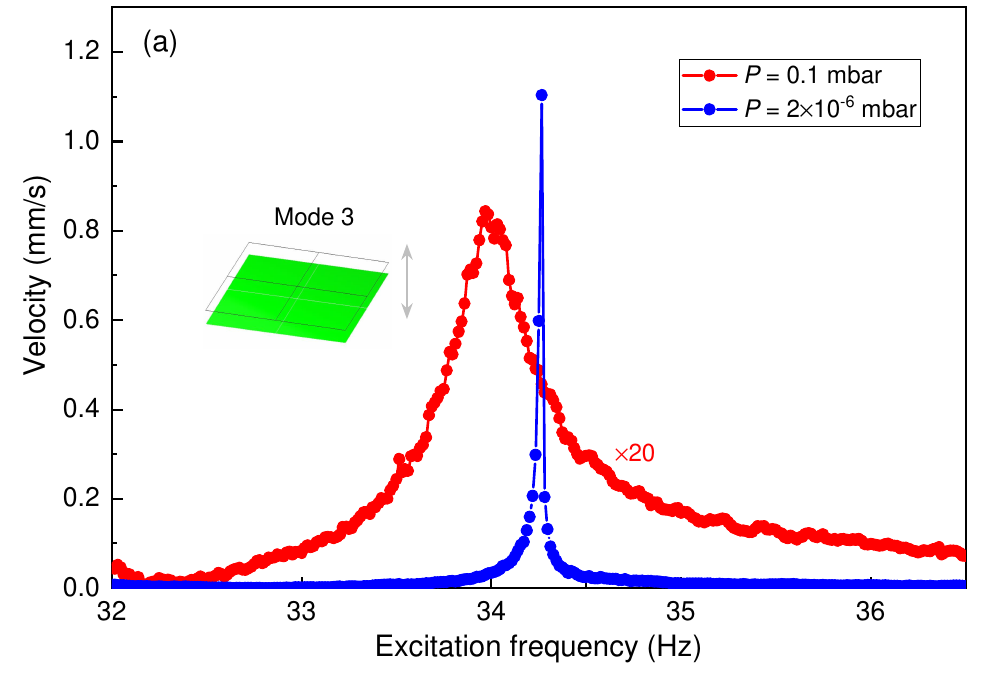}
	\includegraphics[height=5cm]{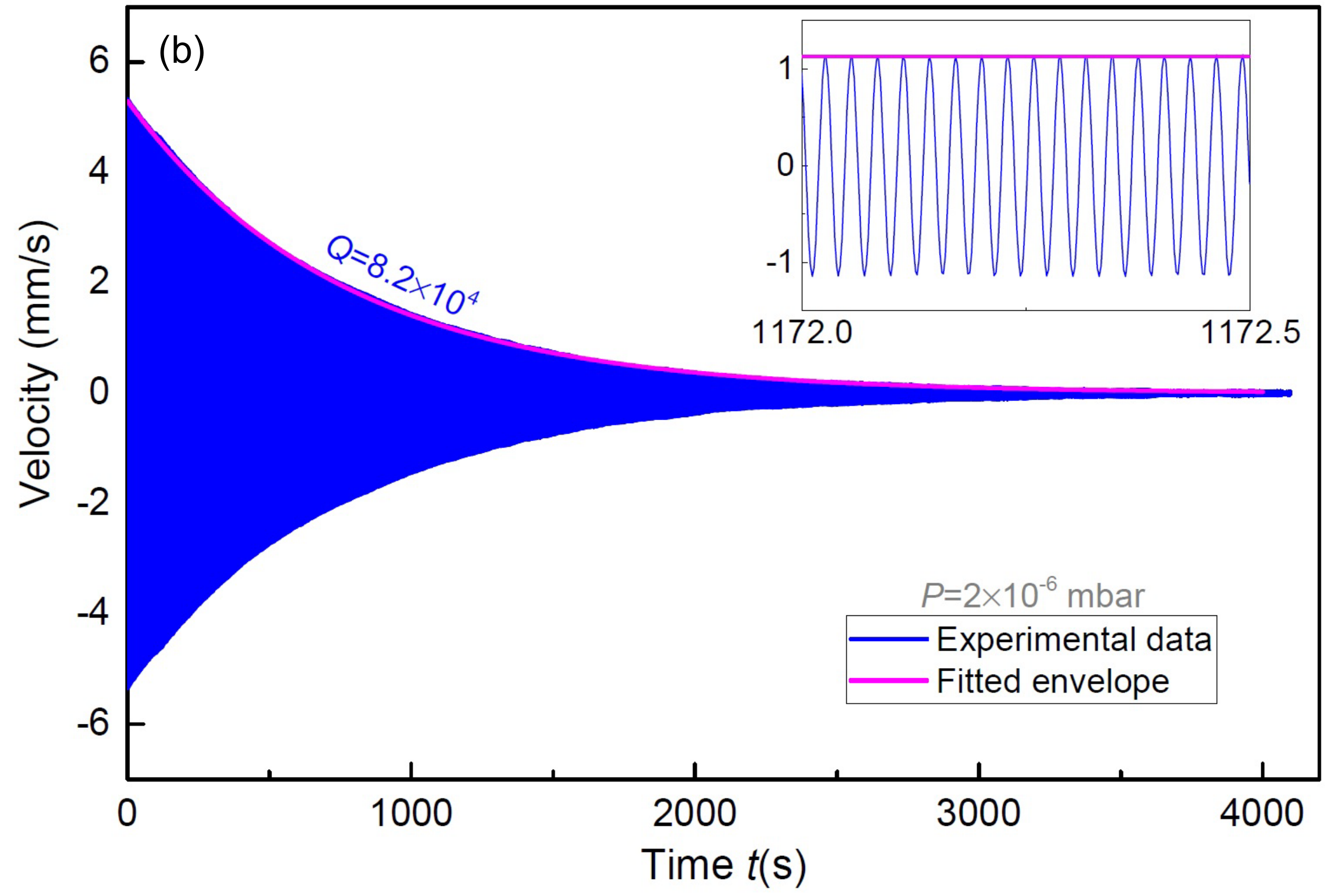}
	\includegraphics[height=5.1cm]{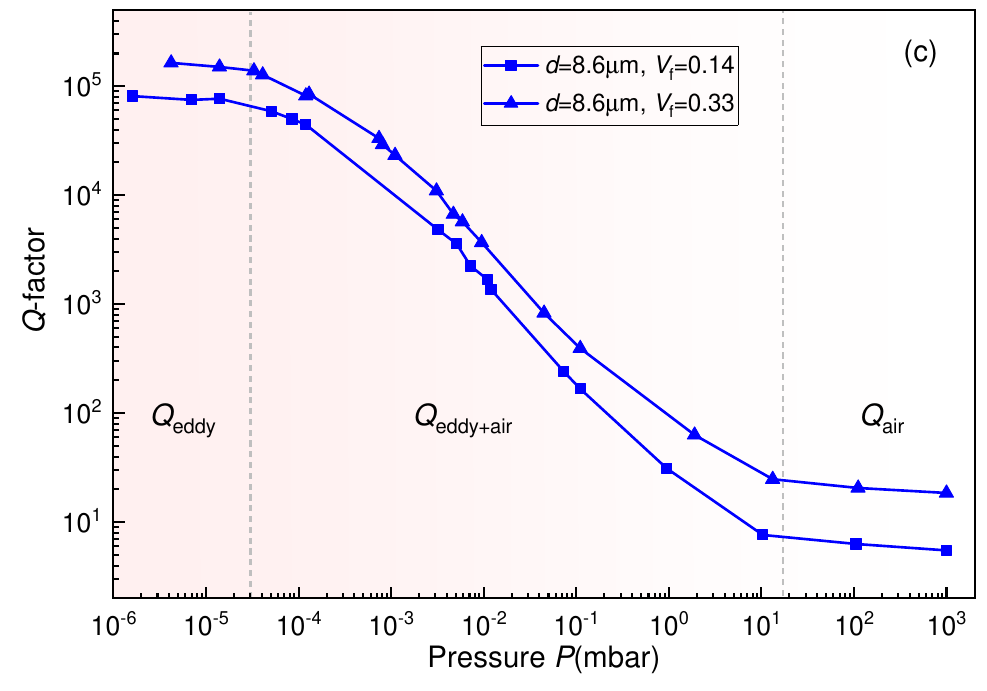}
	\includegraphics[height=5.1cm]{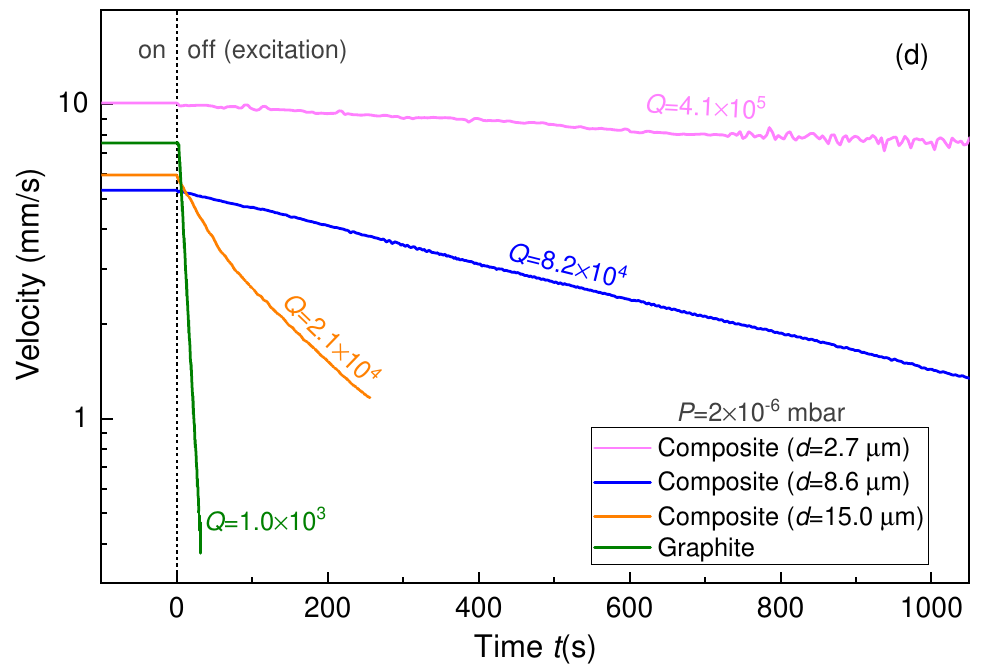}
    \caption{\label{fig:Ringdown measurement and pressure dependent Q} Energy dissipation measurements. (a) Frequency response curves for the translational mode of the $1.8\times1.8\times\SI{0.09}{mm^3}$ levitating composite plate measured at \SI{0.1}{mbar} and \SI{2}{\times 10^{-6} mbar}. The frequency response curve at \SI{0.1}{mbar} has been multiplied by a factor of 20 for visibility. (b) Undriven ringdown of the same composite plate for a duration of \SI{4,000}{s} at \SI{2}{\times 10^{-6} mbar} and its fitted envelope. The time signal for a \SI{0.5}{s} interval is also shown in the inset. (c) The $Q$ as a function of pressure for two sizes of composite plate shows three characteristic regions comprising the region where air damping is dominant (right), region where both air and eddy current damping contribute to dissipation (middle), and region where eddy current damping is dominant (left). (d) Ringdowns of a levitating graphite plate and three composite plates composed of different particle size, revealing that decreasing the particle size results in higher $Q$ of the samples. The dashed line separates the time span between the excitation is on and off.}
\end{figure*}

Since eddy current and air damping \cite{chen2020rigid} are the major sources of dissipation in diamagnetically levitating objects, we minimize the effect of air damping by operating the composite plate resonator in high vacuum ($10^{-6} \si{mbar}$). In Fig. \ref{fig:Ringdown measurement and pressure dependent Q}a, we compare the resonant response of the plate's translational mode in low ($0.1 \si{mbar}$) and high ($10^{-6} \si{mbar}$) vacuum environments. We find an increase in the resonance frequency which we attribute to a reduction in mass loading by the surrounding gas. Moreover, the high vacuum results in a much sharper peak, with much higher $Q$, due to the reduction of air damping effects. In fact, the $Q$ is so high that it is difficult to accurately determine it using a frequency response measurement, due to the limited resolution bandwidth of the measurement setup.

To determine the $Q$ more accurately while also minimizing the influence of spectral broadening, we perform ringdown measurements. These are conducted by first electrostatically exciting the composite plate at its resonance frequency, then switching off the excitation voltage and recording the free vibration decay. The amplitude of the underdamped vibration decays proportional to $\propto e^{-\frac{t}{\tau}}$, where $\tau= \frac{Q}{\pi f_\mathrm{res}}$ is the decay constant and $f_\mathrm{res}$ is the resonance frequency of the plate. In Fig. \ref{fig:Ringdown measurement and pressure dependent Q}b we show a typical measurement for the translational mode of the levitating composite. Note that a very long vibration lifetime of $\sim 4,000$s is observed, corresponding to a $Q$ of $8.2\times 10^{4}$.  In the inset of Fig. \ref{fig:Ringdown measurement and pressure dependent Q}b we also show the free vibrations of the plate over a \SI{0.5}{s} time interval, demonstrating a clear sinusoidal response during the energy decay measurements. It is noted that due to the presence of low-frequency perturbations from the vacuum pump and the environment, the amplitude of the high-$Q$ composites might fluctuate during the ringdown measurements as shown in Fig. \ref{fig:Ringdown measurement and pressure dependent Q}d. However, the fluctuations do not influence the $Q$ factor measurements as they are very small compared to the vibration amplitude.

To ensure that the energy decay constant $\tau$ is not limited by air damping, we sweep the pressure from $10^{-6}- 10^{3}$ mbar and measured $Q$ as a function of air pressure (see Fig. \ref{fig:Ringdown measurement and pressure dependent Q}c). The data for two composite plates with $d=\SI{8.6}{\um}$ particle size and different graphite volume fractions show three distinct regions in the $Q$ versus pressure plot.  When the pressure is reduced below \SI{3e-5}{mbar}, $Q$ reaches a plateau, as shown in Fig. \ref{fig:Ringdown measurement and pressure dependent Q}c. This suggests that air damping has become negligible, and $Q$ is solely limited by eddy currents. The $Q$s shown in the rest of this work are measured at a pressure below \SI{5e-6}{mbar} to eliminate the effect of air damping in our measurements.

\subsection{Tailoring composite properties to suppress eddy currents}
To investigate the effect of the graphite particle size and volume fraction on the levitation forces and the eddy current damping, we fabricate square plates with different graphite volume fractions $V_\mathrm{f}$, side length $L$ and particle size $d$. In Fig. \ref{fig:Ringdown measurement and pressure dependent Q}d we compare the ringdown response of three $1.8\times1.8\times\SI{0.09}{mm^3}$ graphite composite plates with different particle sizes, namely $d=15.0, 8.6$ and $\SI{2.7}{\um}$. We find that the plate that encompasses the smallest particle size exhibits the largest value of $Q$. Remarkably, we observe an increase of nearly $410$ times in $Q$ for the $\SI{2.7}{\um}$ particle composite plates compared to the levitating graphite plate. 

%Recently, it was theoretically shown that diamagnetically levitating graphite resonators can feature record-high Qs upon size reduction \cite{chen2020rigid}. To verify this theory,

%To further investigate the sensitivity of damping to graphite volume fraction, we also fabricate composites with $V_\mathrm{f}$ ranging from 0.14 to 0.45 for two graphite particle sizes namely $d=\SI{8.6}{\um}$ and $d=\SI{17.6}{\um}$

%that the amplitude drops faster with smaller Qs, and the vibration energy of a higher-Q resonator can preserve in the system for a longer time. 

%measure the Qs under different pressures. Fig. \ref{fig:Ringdown measurement and pressure dependent Q}c shows the pressure-dependent Qs for two $1.8\times1.8\times\SI{0.09}{mm^3}$ plates with $d=\SI{8.6}{\um}$ particles but different volume fractions. 

%\newpage
%\subsection{Particle size, volume fraction and plate size-dependent Q}
%%
\begin{figure*}
    \centering
	\includegraphics[height=5.05cm]{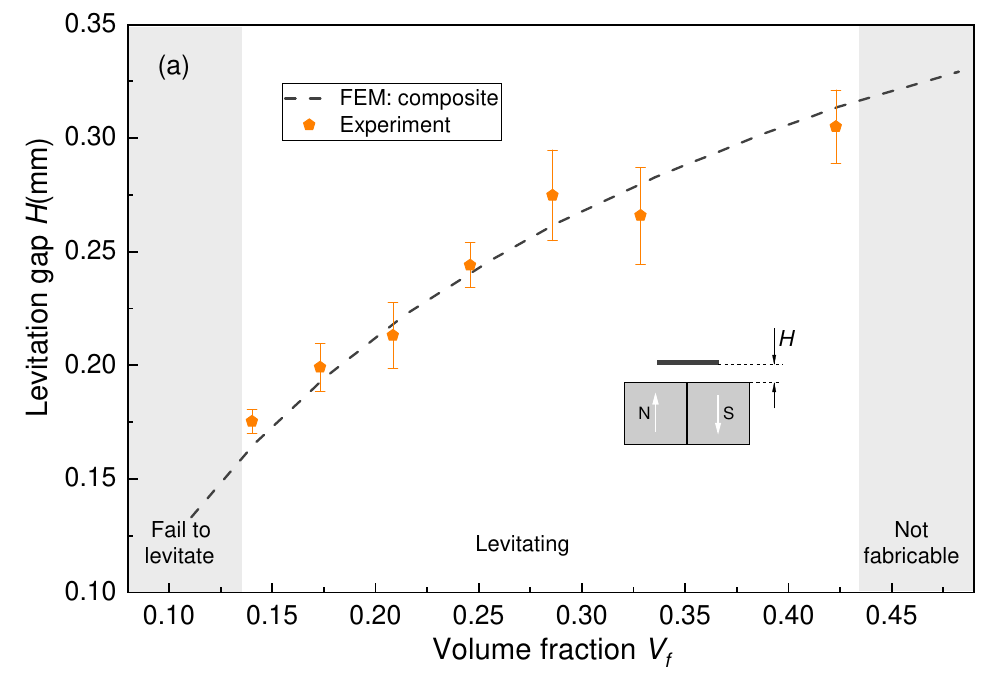}
	\hspace{0.1 cm}
	\includegraphics[height=5cm]{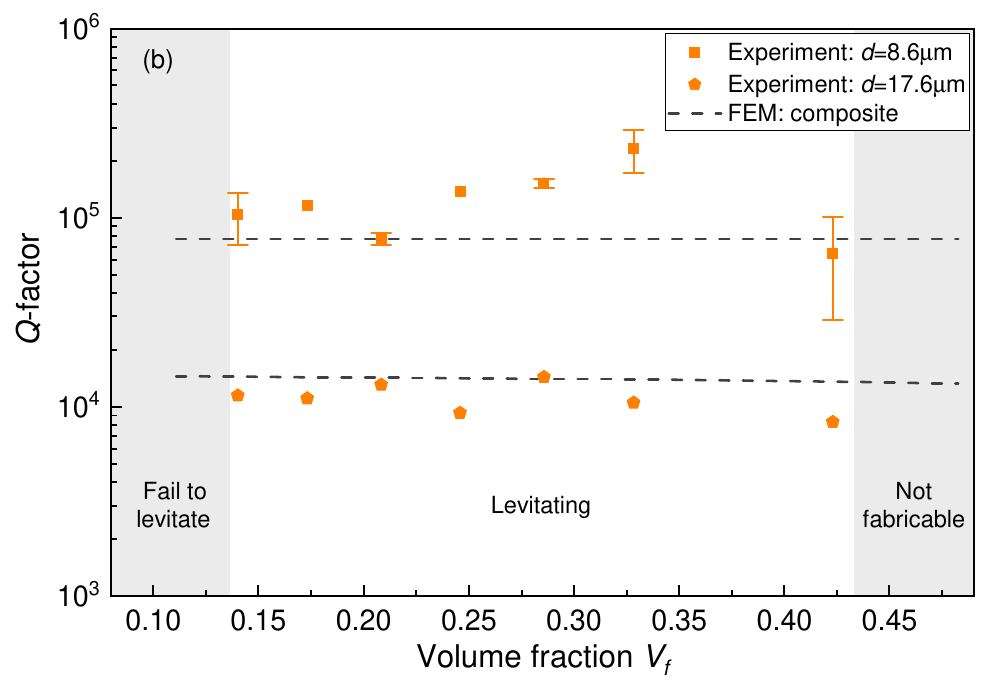}
	\includegraphics[height=5cm]{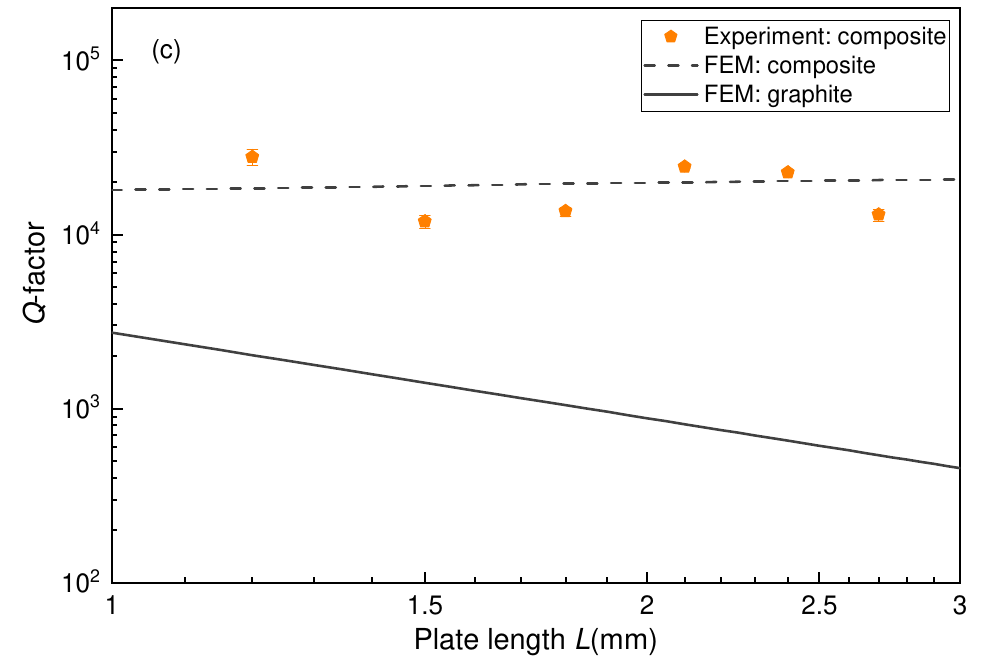}
	\includegraphics[height=5cm]{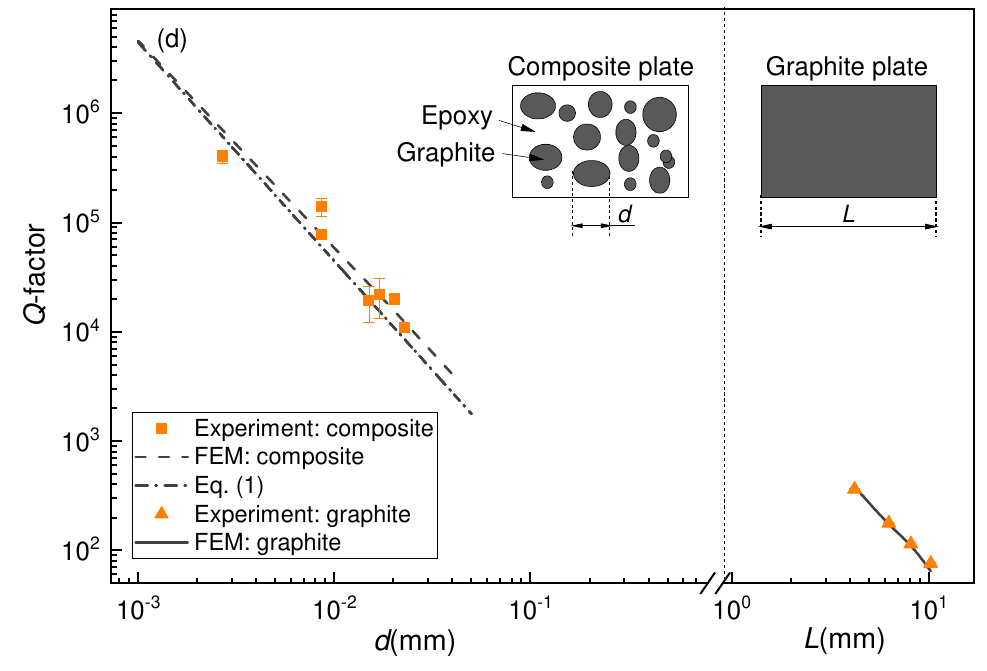}
    \caption{\label{fig:Q-composite} Dependence of the levitation force and dissipation on composite particle and plate size. (a) Levitation gap $H$ of the $1.8\times1.8\times\SI{0.09}{mm^3}$ plate with \SI{17.6}{\um} particles as a function of volume fraction. (b) $Q$-factor of two $1.8\times1.8\times\SI{0.09}{mm^3}$ plates with different particle size as a function of volume fraction. The gray area in Fig. \ref{fig:Q-composite}a-b represents the volume fraction of which the composites can not be levitated. (c) $Q$-factor of square composite plates with a thickness of \SI{90}{\um} as a function of plate side length $L$. The composite plates are made from $d$=\SI{17.6}{\um} particles with 0.21 volume fraction.
    (d) Dependence of $Q$ on particle size. The left side of the graph (before the short dashed line) shows the $Q$ of composite plates with varying graphite particle size. Since the $Q$ is only weakly dependent on the volume fraction and side length (see Fig. \ref{fig:Q-composite}b-c), the error bars in the data on the left side of Fig. 4d are obtained by analyzing the $Q$s obtained from plates with a thickness of \SI{90}{\um} but different side lengths (1.2-\SI{2.7}{\mm}) and volume fraction (0.14-0.32) at fixed $d$. The right side of the graph (after the short dashed line) shows the experimental $Q$ of levitating plates made of pyrolytic graphite with \SI{0.28}{mm} thickness and different side lengths $L$ on the $x$-axis. The insets show schematics of the composite and graphite plate. In Fig. 4, the dashed and solid lines correspond to the FEM simulations for composite and graphite plates as described in S3, respectively. Moreover, the dashed-dotted line in Fig. 4d represents the $Q$s obtained from Eq. (1). The dots represent experimental data.}
\end{figure*}

To better understand this observation and gain deeper insight into the mechanisms accountable for $Q$ enhancement, simulations based on Finite Element Method (FEM) are performed to calculate the levitation height and eddy current damping force using COMSOL multiphysics. These simulations are carried out assuming that the graphite particles have a spherical shape and are distributed inside the matrix (see S3 for details of the numerical modelling and parameter values used in our simulations). We note that graphite is inherently anisotropic \cite{simon2000diamagnetic}. However, in our fabrication procedure graphite particles are randomly oriented in the epoxy matrix, and thus by considering all possible orientations in the matrix, the local anisotropy can be averaged out and the effective macroscopic behavior can be viewed isotropic. For this reason, in our study we treat the magnetic susceptibility of graphite as an effective value $\chi_\mathrm{eff}$ which we evaluate by fitting our FEM simulations to the measured levitation height of the composite from experiments (see Fig. S8 for more details).

In Fig. \ref{fig:Q-composite}a we show the levitation height of the composite plates with particle size $d=\SI{17.6}{\um}$ as a function of volume fraction $V_\mathrm{f}$. We find that composites with a graphite volume fraction below 14\% ($V_\mathrm{f}<0.14$) do not provide sufficient diamagnetic force to counteract gravity and thus do not levitate. For composites with a graphite volume fraction above 43\% ($V_\mathrm{f}>0.43$), the samples can not be produced with sufficient structural integrity due to the high particle content. Between these two limits, we observe a steady increase in the levitation height which agrees well with the simulations. These results indicate that the increase of the magnetic force is dominant over the increase in the overall gravitational force through the higher mass density of the graphite particles compared to the epoxy, see Table II in S3.

% In order to see how the $V_\mathrm{f}$ influences the levitation of the composites, we first measure their levitation height as a function of $V_\mathrm{f}$.
 
We also study the influence of volume fraction $V_\mathrm{f}$ on the measured $Q$ for composite plates with particle sizes $d=\SI{8.6}{\um}$ and $d=\SI{17.6}{\um}$, as shown in Fig. \ref{fig:Q-composite}b. It is interesting to see that unlike levitation height, $Q$ does not significantly change with $V_\mathrm{f}$, even though the measured  bulk conductivity reveals an increase in the conductivity with the increase of $V_\mathrm{f}$ (see S4). This result suggests that
the variations in bulk conductivity do not contribute considerably to the observed dissipation. A similar effect is seen in Fig. \ref{fig:Q-composite}c, where we show the experimentally obtained $Q$ of square plates with different side lengths $L$, that are cut out of the same composite with $d=\SI{17.6}{\um}$ and $V_\mathrm{f}=0.21$. It is observed from the figure that a reduction in side length does not substantially increase $Q$. This observation contrasts with $Q$s estimated from COMSOL simulations for pyrolytic graphite plates, that increase close to an order of magnitude with reducing $L$\cite{chen2020rigid}.

The volume fraction and plate size independent $Q$s obtained from both experiments and simulations in Fig. \ref{fig:Q-composite}b-c indicate that the majority of eddy current damping occurs inside the graphite particles and is not caused by currents flowing in between them. Thus, increasing the particle density increases the stored kinetic energy $E_\mathrm{k}$ (proportional to the mass) by approximately the same factor as the eddy current dissipation $E_\mathrm{d}$ (proportional to number of particles) (see Fig. S9 in S3.2), such that the $Q$, which is proportional to their ratio $E_\mathrm{k}/E_\mathrm{d}$, remains nominally constant.

After having investigated the effect of volume fraction and composite plate size, we now investigate the effect of graphite particle size $d$ on $Q$.
%The reason behind this is similar with volume fraction. With the increase of $L$, damping force and mass both increase (see Fig. S5d-f), thus canceling each other in Eq. \ref{eq: Q}. Therefore, from the results shown in Fig. \ref{fig:Q-composite}b,c, it can be concluded
It can be observed from both the experimental and numerical results (see left side of Fig. \ref{fig:Q-composite}d) that reducing the particle size $d$ results in a clear increase in the $Q$ of the composite plates. The $Q$ increases from about 10,000 at $d$=\SI{22.7}{\um}, to a value as high as 460,000 at $d=\SI{2.7}{\um}$, which is to our knowledge a record value for passively levitating diamagnetic resonators at room temperature. On the right side of Fig. \ref{fig:Q-composite}d the $Q$s of pyrolytic graphite plates with varying side lengths are shown. These plates also show an increasing $Q$ with decreasing plate size\cite{chen2020rigid}.

%Best first present all experimental results in Fig. 4b-d and only then discuss the simulations/analytical model.
%The FEM simulation methodology is outlined in SIx (see Fig. S5g-h) and results in the dashed and solid lines in Figs. \ref{fig:Q-composite}b,c where the parameters y=10 m/s and z=20 N are used as fit parameters to obtain agreement with experimental values. 

\section{Discussion and conclusions}
%\subsection{Analytical model}
To understand these findings, and in particular the increase in $Q$ as a function of $d$, we use Faraday's law and obtain an analytic expression for the $Q$ of a graphite/epoxy composite plate that moves in a magnetic field (see S5 for the detailed derivation):
\begin{equation}
    \label{eq: SI-Q simple model for composite}
    Q=\frac{80\pi f_\mathrm{res}\rho_\mathrm{r}((\rho_\mathrm{g}-\rho_\mathrm{e})+\rho_\mathrm{e}/V_\mathrm{f})}{(C_\mathrm{r}d)^2\nabla^2 B},
\end{equation}
where $\rho_\mathrm{r}$ is the resistivity, $\rho_\mathrm{g}$ is the density of graphite, $\rho_\mathrm{e}$ is the density of epoxy, and $C_\mathrm{r}$ is the effective particle size factor which we use to account for experimental deviations from the theoretical model due to variations in particle size, composition, morphology and distribution. Moreover, $\nabla^2 B$ represents the Laplacian of the magnetic field, which is
\begin{equation}
    \nabla^2 B=\frac{\int_{V_\mathrm{plate}}\left(\frac{\mathrm{d}B}{\mathrm{d}z}\right)^2\mathrm{d}V_\mathrm{plate}}{V_\mathrm{plate}}.
\end{equation}
 To compare our experimental findings in Fig. \ref{fig:Q-composite}d to the analytical expression Eq. (\ref{eq: SI-Q simple model for composite}), we take $f_\mathrm{res}=\SI{35}{Hz}, \rho_\mathrm{r}=\SI{5e-6}{\ohm\cdot m}, \rho_\mathrm{g}=\SI{2260}{kg/m^3}, \rho_\mathrm{e}=\SI{1100}{kg/m^3}$ , and use COMSOL simulations to calculate $\nabla^2 B=\SI{1.1e+06}{(T/m)^2}$ for a $1.8\times1.8\times\SI{0.09}{mm^3}$ plate that levitates \SI{0.26}{mm} above the magnets, corresponding to a composite with $V_\mathrm{f}=0.32$ (see Fig. \ref{fig:Q-composite}a). Using these values and $C_\mathrm{r}=6.3$ as a fit parameter, we can match the experimental data shown in Fig. \ref{fig:Q-composite}d with good accuracy. These results show that the $Q$ in our levitating composites is inversely proportional to $d^2$, providing evidence that the strong dependence of $Q$ on particle size can be mainly accounted for using Eq. \eqref{eq: SI-Q simple model for composite}, which is based on dissipation due to eddy currents that flow inside the graphite particles. The high sensitivity of $Q$ to $d$, allows us to  engineer and increase the $Q$ of our levitating resonators by using different particle size while keeping the macroscopic dimensions of the plate constant. The highest $Q$ we obtain with this fabrication process is $4.6\times10^5$ for a $2.7\times2.7\times\SI{0.09}{m^3}$ composite plate with $d=\SI{2.7}{\um}$ particles and $V_\mathrm{f}=0.21$ volume fraction, which is two orders of magnitude higher than a pyrolytic graphite plate of the same size as shown in Fig. \ref{fig:Q-composite}d.

%\subsection{Discussion}
It is of interest to note that extrapolation of the graphite plate data in the right side of Fig. \ref{fig:Q-composite}d to smaller values of $d$ leads to much higher values of $Q$ than that are obtained experimentally with the composites in the left part of Fig. \ref{fig:Q-composite}d. Several mechanisms might account for this difference, including the random orientation of the graphite particles in the composite, the particle size and shape variations,  inactive layers on the particle surfaces and material parameter differences between the graphite in the plates and particles. In Fig. \ref{fig:Q-composite}d the combined effect of these mechanisms are captured by the effective particle size factor $C_\mathrm{r}$. Although we can not fully account quantitatively for the relatively large value of $C_\mathrm{r}=6.3$ of this factor, possibly a small fraction of larger particles or clusters of particles in the composite accounts for a large part of the damping force.  Microscopic images of the composites in S2 support this hypothesis, by showing that the dispersion of the particles is random and less homogeneous inside the epoxy matrix with local particle clusters. It might also be that not all sources of damping are included in Eq. (\ref{eq: SI-Q simple model for composite}) and more sophisticated models will need to be developed. Nevertheless, we foresee that by further control of the particle size and optimization of its distribution, levitating composites can achieve $Q$s above 1 million for millimeter composites with \SI{1}{\um} or smaller particles.

\begin{figure*}
    \centering
	\includegraphics[height=5cm]{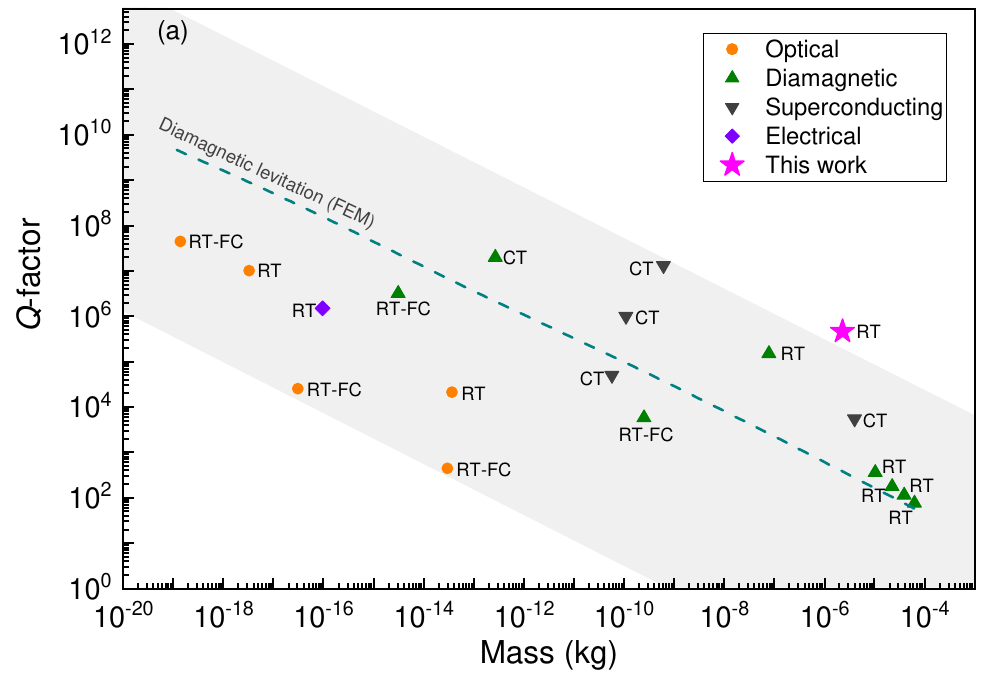}
	\includegraphics[height=5cm]{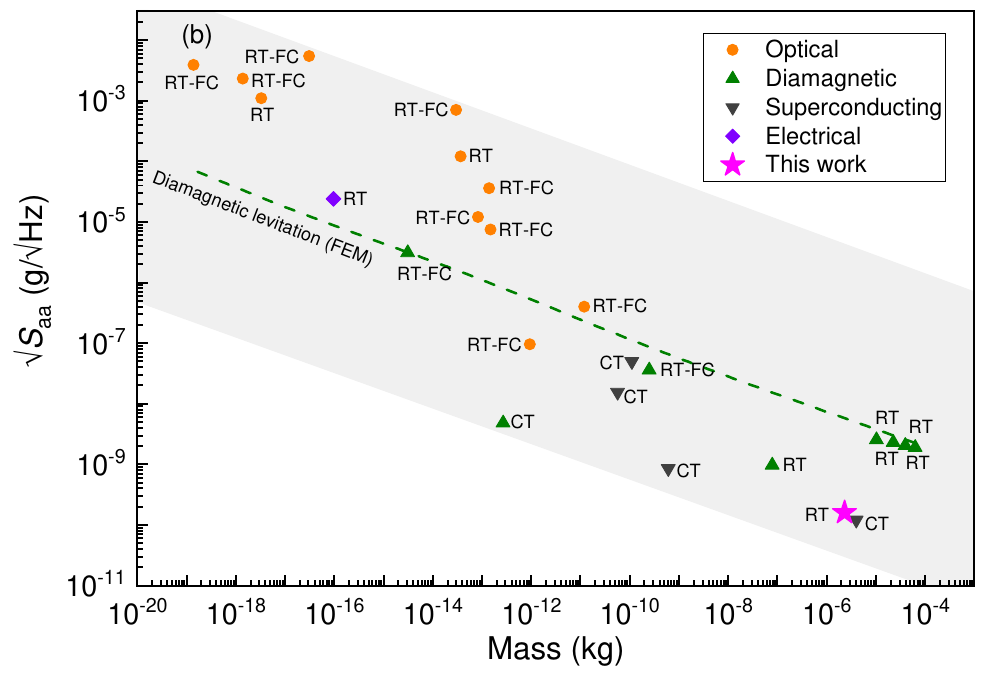}
    \caption{\label{fig:sensitivity} (a) $Q$-factor versus mass of different levitating systems (optical \cite{vovrosh2017parametric,gieseler2012subkelvin,li2011millikelvin,ranjit2016zeptonewton,ranjit2015attonewton}, diamagnetic \cite{lewandowski2021high,leng2021mechanical,PhysRevApplied.16.L011003,chen2020rigid,slezak2018cooling}, superconducting \cite{vinante2020ultralow,gieseler2020single,wang2019dynamics,timberlake2019acceleration} and electrical \cite{pontin2020ultranarrow}). (b) Plot of acceleration noise floor against mass of different levitating accelerometers (optical \cite{gieseler2012subkelvin,li2011millikelvin,vovrosh2017parametric,ranjit2016zeptonewton,ranjit2015attonewton,hempston2017force,kawasaki2020high,moore2014search,rider2018single,monteiro2017optical,monteiro2020force}, diamagnetic \cite{leng2021mechanical,PhysRevApplied.16.L011003,lewandowski2021high,chen2020rigid,slezak2018cooling}, superconducting \cite{vinante2020ultralow,gieseler2020single,wang2019dynamics,timberlake2019acceleration} and electrical \cite{pontin2020ultranarrow}).  RT stands for $Q$s measured at room temperature without feedback cooling, CT stands for $Q$s measured at cryogenic temperature without feedback cooling, RT-FC stands for natural $Q$s estimated from feedback cooling measurements. The $Q$ and $\sqrt{S_\mathrm{aa}}$ of different levitating systems are also shown in Table III-V of S6. The $Q$ and $\sqrt{S_\mathrm{aa}}$ shown as dashed lines are simulated using COMSOL for graphite plates with different size $L$ (detailed material parameter values used for these simulations can be found in Table 1 of S3). In the simulations the plate thickness $t$ and magnet size $D$ are taken proportional to the plate side length ($D=1.2 L$ and $t=0.03 L$). The grey area in Fig. \ref{fig:sensitivity}a-b sets the boundary of the $Q$ and acceleration noise floor against mass of available levitodynamic systems in the literature.}
\end{figure*}
%\subsection{Limiting source of $Q$-factor in levitating composites}

%\subsection{$Q$-factors of levitation systems}

The combination of high $Q$ and large mass of the levitating composites promises low noise floor levels in accelerometry. In Fig. \ref{fig:sensitivity} we benchmark the presented levitating composite plates against state-of-the-art levitodynamic systems by plotting mass against $Q$ (Fig. \ref{fig:sensitivity}a) and the square root of the acceleration noise power spectral density $S_{aa} \propto f_\mathrm{res}/(m Q)$ \cite{timberlake2019acceleration} (Fig. \ref{fig:sensitivity}b), which is a measure of the limit of detection of an accelerometer. The plots compare a range of superconducting, diamagnetically, electrically, and optically levitating systems, at room temperature (labeled with RT in Fig. 5),  at cryogenic temperature (CT) or using feedback cooling (RT-FC). Note that RT-FC stands for natural $Q$s that are estimated from feedback cooling measurements. The plots also show the theoretical estimates of $Q$ and $\sqrt{S_{aa}}$ (dashed lines) as a function of mass for diamagnetically levitating pyrolytic graphite.  
It appears from this benchmark that in terms of acceleration noise floor and $Q$, diamagnetic composites stand out, providing the possibility to levitate large, high-Q objects using relatively weak fields from permanent magnets. The combination of large levitating proof mass and high $Q$ make these composites attractive materials for realizing next generation room temperature accelerometers with theoretical sensitivities as low as $\SI{0.16}{ng/\sqrt{Hz}}$, that are comparable to superconducting levitodynamic systems at cryogenic temperatures (Fig. \ref{fig:sensitivity}b).

%\subsection{Conclusion} 
In conclusion, we demonstrate diamagnetic high $Q$ composite plate resonators consisting of graphite particles dispersed in an epoxy matrix that can be levitated at room temperature above permanent magnets with graphite volume fractions as low as 14\%. By insulating the graphite particles, eddy currents are reduced and confined within the particles, allowing us to suppress the associated damping forces. This enables a remarkable enhancement in $Q$, reaching values as high as nearly 0.5 million at room temperature. Measurements of the dependence of damping to particle volume fraction, plate length, and particle size are compared to FEM models, and show good agreement with an analytical model for eddy current damping forces that predicts $Q$ to be inversely proportional to the squared particle size $Q\propto 1/d^2$. 
%This is new for me. Don't bring new information at the end of the paper. Using the perfectly isolated spherical particle model, the $Q$s of the composite plates were anticipated to be $40 \times$ higher than the experimental values. In our study, we attributed the discrepancy to the effective particle size due to heterogeneity in the particle size and distribution. Our fabrication platform enables the first steps to gain deeper understanding of complex eddy current formation and conductive pathways as particles start to percolate. 
Reduction of the particle size and optimization of particle distribution and orientation, can lead to novel composites that further enhance the performance of future macroscopic levitating devices used as accelerometers\cite{timberlake2019acceleration}, gravimeters\cite{middlemiss2016measurement,schmole2016micromechanical,marletto2017gravitationally,belenchia2018quantum}, or sensors for exploring macroscopic limits of quantum mechanics \cite{whittle2021approaching,michimura2020quantum,yu2020quantum,catano2020high}.

\section{Methods}
\subsection{Composite fabrication}
 Graphite micro powders (purity $>$99.9\%) with mean sizes from $2.7-\SI{22.7}{\um}$ are purchased from Nanografi Nano Technology. Particle size distribution measurements are performed using Malvern Mastersizer 3000 on {0.1}\% weight/volume aqueous solution of the powders using sodium dodecyl sulfate solution as surfactant. The particle size $d$ of each type of powder is represented by the mean value of the distribution. The morphology of these powders is confirmed via Scanning Electron Microscopy (JEOL JSM-7500F).
 
The details of the graphite composite fabrication process are shown in Fig. S1. First, the two components of the epoxy (Epotek 302-3M from Gentec Benelux) are mixed at 3500 rpm for 5 minutes in a  Dual Asymmetric Centrifuge mixer (DAC 150.1 FVZ-K ) followed by the addition and mixing of the  graphite powder at 500 rpm for 5 minutes. To reduce the viscosity of the resulting graphite-epoxy paste, ethanol is added, and further mixed at 500 rpm for \SI{5}{mins}. This maximises dispersion and homogeneity of the paste with the graphite particles in the epoxy-ethanol matrix. The paste is then transferred into circular holes ($\phi=\SI{10}{mm}$) in a thin plastic mould with thickness of \SI{0.12}{mm} on the top of a flat steel mould. The deposited paste is left at room temperature and pressure for 30 minutes to let the ethanol fully evaporate before curing the epoxy in order to minimise porosity. The graphite/epoxy paste is then compressed by steel moulds and cured in an oven at \SI{100}{\degree C} for around \SI{12}{hours}. After curing, an Optec micro laser cutter is used to cut the composite into square plates with desired lengths. Finally, fine sand paper (\SI{5}{\um} grain) is used to polish composite surface to the desired thickness.

\subsection{Measurement}
In our experiments, the excitation voltage is generated by the Polytec MSA400 vibrometer for the resonance frequency measurements, and by a function generator for the ringdown response measurements. 
The electrostatic force is generated as shown in Fig. 2a, by applying a voltage difference between the magnets beneath the levitating plate. To isolate the magnets from one another we use Kapton tape. When a voltage is applied between the two electrodes, the levitating plate acts as a floating electrode between the two electrodes, thereby forming a capacitive divider. In the area at which the plate overlaps with the electrodes, an electrostatic downward force is exerted
that depends on the overlap area, voltage difference, and gap size. Since the electrostatic force is proportional to the square of the voltage, a DC offset voltage is added to make sure the electrostatic force has a component of the same frequency as the output voltage.  Finally, to read out the motion, a Polytec LDV is used. The LDV measurements are conducted in a vacuum chamber over a pressure range of $10^{-6}-\SI{1000}{mbar}$ at room temperature.

\section*{Acknowledgments}
%This work is supported by the European Union’s Horizon 2020 research and innovation program under Grant Agreements No. 802093 (ERC starting grant ENIGMA) and 881603 (Graphene Flagship). The work has also received funding from EMPIR program co-financed by the Participating States (785219).
This work was carried out under the 17FUN05 PhotOQuanT project, which has received funding from the EMPIR program, co-financed by the Participating States and the European Union’s Horizon 2020 research and innovation program. This work has also received funding from ERC starting grant ENIGMA (802093) and Graphene Flagship (881603). X.C acknowledges financial support from China Scholarship Council.

%\nocite{*}

\bibliography{ref-high-Q}% Produces the bibliography via BibTeX.

\begin{thebibliography}{44}%
\makeatletter
\providecommand \@ifxundefined [1]{%
 \@ifx{#1\undefined}
}%
\providecommand \@ifnum [1]{%
 \ifnum #1\expandafter \@firstoftwo
 \else \expandafter \@secondoftwo
 \fi
}%
\providecommand \@ifx [1]{%
 \ifx #1\expandafter \@firstoftwo
 \else \expandafter \@secondoftwo
 \fi
}%
\providecommand \natexlab [1]{#1}%
\providecommand \enquote  [1]{``#1''}%
\providecommand \bibnamefont  [1]{#1}%
\providecommand \bibfnamefont [1]{#1}%
\providecommand \citenamefont [1]{#1}%
\providecommand \href@noop [0]{\@secondoftwo}%
\providecommand \href [0]{\begingroup \@sanitize@url \@href}%
\providecommand \@href[1]{\@@startlink{#1}\@@href}%
\providecommand \@@href[1]{\endgroup#1\@@endlink}%
\providecommand \@sanitize@url [0]{\catcode `\\12\catcode `\$12\catcode
  `\&12\catcode `\#12\catcode `\^12\catcode `\_12\catcode `\%12\relax}%
\providecommand \@@startlink[1]{}%
\providecommand \@@endlink[0]{}%
\providecommand \url  [0]{\begingroup\@sanitize@url \@url }%
\providecommand \@url [1]{\endgroup\@href {#1}{\urlprefix }}%
\providecommand \urlprefix  [0]{URL }%
\providecommand \Eprint [0]{\href }%
\providecommand \doibase [0]{https://doi.org/}%
\providecommand \selectlanguage [0]{\@gobble}%
\providecommand \bibinfo  [0]{\@secondoftwo}%
\providecommand \bibfield  [0]{\@secondoftwo}%
\providecommand \translation [1]{[#1]}%
\providecommand \BibitemOpen [0]{}%
\providecommand \bibitemStop [0]{}%
\providecommand \bibitemNoStop [0]{.\EOS\space}%
\providecommand \EOS [0]{\spacefactor3000\relax}%
\providecommand \BibitemShut  [1]{\csname bibitem#1\endcsname}%
\let\auto@bib@innerbib\@empty
%</preamble>
\bibitem [{\citenamefont {MacCabe}\ \emph {et~al.}(2020)\citenamefont
  {MacCabe}, \citenamefont {Ren}, \citenamefont {Luo}, \citenamefont {Cohen},
  \citenamefont {Zhou}, \citenamefont {Sipahigil}, \citenamefont
  {Mirhosseini},\ and\ \citenamefont {Painter}}]{maccabe2020nano}%
  \BibitemOpen
  \bibfield  {author} {\bibinfo {author} {\bibfnamefont {G.~S.}\ \bibnamefont
  {MacCabe}}, \bibinfo {author} {\bibfnamefont {H.}~\bibnamefont {Ren}},
  \bibinfo {author} {\bibfnamefont {J.}~\bibnamefont {Luo}}, \bibinfo {author}
  {\bibfnamefont {J.~D.}\ \bibnamefont {Cohen}}, \bibinfo {author}
  {\bibfnamefont {H.}~\bibnamefont {Zhou}}, \bibinfo {author} {\bibfnamefont
  {A.}~\bibnamefont {Sipahigil}}, \bibinfo {author} {\bibfnamefont
  {M.}~\bibnamefont {Mirhosseini}},\ and\ \bibinfo {author} {\bibfnamefont
  {O.}~\bibnamefont {Painter}},\ }\bibfield  {title} {\bibinfo {title}
  {Nano-acoustic resonator with ultralong phonon lifetime},\ }\href@noop {}
  {\bibfield  {journal} {\bibinfo  {journal} {Science}\ }\textbf {\bibinfo
  {volume} {370}},\ \bibinfo {pages} {840} (\bibinfo {year}
  {2020})}\BibitemShut {NoStop}%
\bibitem [{\citenamefont {Bereyhi}\ \emph {et~al.}(2022)\citenamefont
  {Bereyhi}, \citenamefont {Beccari}, \citenamefont {Groth}, \citenamefont
  {Fedorov}, \citenamefont {Arabmoheghi}, \citenamefont {Engelsen},\ and\
  \citenamefont {Kippenberg}}]{beccari2103hierarchical}%
  \BibitemOpen
  \bibfield  {author} {\bibinfo {author} {\bibfnamefont {M.}~\bibnamefont
  {Bereyhi}}, \bibinfo {author} {\bibfnamefont {A.}~\bibnamefont {Beccari}},
  \bibinfo {author} {\bibfnamefont {R.}~\bibnamefont {Groth}}, \bibinfo
  {author} {\bibfnamefont {S.}~\bibnamefont {Fedorov}}, \bibinfo {author}
  {\bibfnamefont {A.}~\bibnamefont {Arabmoheghi}}, \bibinfo {author}
  {\bibfnamefont {N.}~\bibnamefont {Engelsen}},\ and\ \bibinfo {author}
  {\bibfnamefont {T.}~\bibnamefont {Kippenberg}},\ }\bibfield  {title}
  {\bibinfo {title} {Hierarchical tensile structures with ultralow mechanical
  dissipation},\ }\href@noop {} {\bibfield  {journal} {\bibinfo  {journal}
  {Nature Communications}\ }\textbf {\bibinfo {volume} {13}},\ \bibinfo {pages}
  {3097} (\bibinfo {year} {2022})}\BibitemShut {NoStop}%
\bibitem [{\citenamefont {Kemp}\ \emph {et~al.}(2019)\citenamefont {Kemp},
  \citenamefont {Franzi}, \citenamefont {Haase}, \citenamefont {Jongewaard},
  \citenamefont {Whittaker}, \citenamefont {Kirkpatrick},\ and\ \citenamefont
  {Sparr}}]{kemp2019high}%
  \BibitemOpen
  \bibfield  {author} {\bibinfo {author} {\bibfnamefont {M.~A.}\ \bibnamefont
  {Kemp}}, \bibinfo {author} {\bibfnamefont {M.}~\bibnamefont {Franzi}},
  \bibinfo {author} {\bibfnamefont {A.}~\bibnamefont {Haase}}, \bibinfo
  {author} {\bibfnamefont {E.}~\bibnamefont {Jongewaard}}, \bibinfo {author}
  {\bibfnamefont {M.~T.}\ \bibnamefont {Whittaker}}, \bibinfo {author}
  {\bibfnamefont {M.}~\bibnamefont {Kirkpatrick}},\ and\ \bibinfo {author}
  {\bibfnamefont {R.}~\bibnamefont {Sparr}},\ }\bibfield  {title} {\bibinfo
  {title} {A high q piezoelectric resonator as a portable vlf transmitter},\
  }\href@noop {} {\bibfield  {journal} {\bibinfo  {journal} {Nature
  communications}\ }\textbf {\bibinfo {volume} {10}},\ \bibinfo {pages} {1}
  (\bibinfo {year} {2019})}\BibitemShut {NoStop}%
\bibitem [{\citenamefont {Kumar}\ \emph {et~al.}(2022)\citenamefont {Kumar},
  \citenamefont {Gupta}, \citenamefont {Pitchappa}, \citenamefont {Tan},
  \citenamefont {Chattopadhyay}, \citenamefont {Ducournau}, \citenamefont
  {Wang}, \citenamefont {Chong}, ,\ and\ \citenamefont
  {Singh}}]{kumar2022activ}%
  \BibitemOpen
  \bibfield  {author} {\bibinfo {author} {\bibfnamefont {A.}~\bibnamefont
  {Kumar}}, \bibinfo {author} {\bibfnamefont {M.}~\bibnamefont {Gupta}},
  \bibinfo {author} {\bibfnamefont {P.}~\bibnamefont {Pitchappa}}, \bibinfo
  {author} {\bibfnamefont {T.~C.}\ \bibnamefont {Tan}}, \bibinfo {author}
  {\bibfnamefont {U.}~\bibnamefont {Chattopadhyay}}, \bibinfo {author}
  {\bibfnamefont {G.}~\bibnamefont {Ducournau}}, \bibinfo {author}
  {\bibfnamefont {N.}~\bibnamefont {Wang}}, \bibinfo {author} {\bibfnamefont
  {Y.}~\bibnamefont {Chong}}, ,\ and\ \bibinfo {author} {\bibfnamefont
  {R.}~\bibnamefont {Singh}},\ }\bibfield  {title} {\bibinfo {title} {Active
  ultrahigh-q ($0.2 \times 10^6$) thz topological cavities on a chip},\
  }\href@noop {} {\bibfield  {journal} {\bibinfo  {journal} {Advanced
  Materials}\ }\textbf {\bibinfo {volume} {2202370}},\ \bibinfo {pages} {1}
  (\bibinfo {year} {2022})}\BibitemShut {NoStop}%
\bibitem [{\citenamefont {Cata{\~n}o-Lopez}\ \emph {et~al.}(2020)\citenamefont
  {Cata{\~n}o-Lopez}, \citenamefont {Santiago-Condori}, \citenamefont
  {Edamatsu},\ and\ \citenamefont {Matsumoto}}]{catano2020high}%
  \BibitemOpen
  \bibfield  {author} {\bibinfo {author} {\bibfnamefont {S.~B.}\ \bibnamefont
  {Cata{\~n}o-Lopez}}, \bibinfo {author} {\bibfnamefont {J.~G.}\ \bibnamefont
  {Santiago-Condori}}, \bibinfo {author} {\bibfnamefont {K.}~\bibnamefont
  {Edamatsu}},\ and\ \bibinfo {author} {\bibfnamefont {N.}~\bibnamefont
  {Matsumoto}},\ }\bibfield  {title} {\bibinfo {title} {High-q milligram-scale
  monolithic pendulum for quantum-limited gravity measurements},\ }\href@noop
  {} {\bibfield  {journal} {\bibinfo  {journal} {Physical review letters}\
  }\textbf {\bibinfo {volume} {124}},\ \bibinfo {pages} {221102} (\bibinfo
  {year} {2020})}\BibitemShut {NoStop}%
\bibitem [{\citenamefont {Shin}\ \emph {et~al.}(2022)\citenamefont {Shin},
  \citenamefont {Cupertino}, \citenamefont {de~Jong}, \citenamefont
  {Steeneken}, \citenamefont {Bessa},\ and\ \citenamefont
  {Norte}}]{shin2022spiderweb}%
  \BibitemOpen
  \bibfield  {author} {\bibinfo {author} {\bibfnamefont {D.}~\bibnamefont
  {Shin}}, \bibinfo {author} {\bibfnamefont {A.}~\bibnamefont {Cupertino}},
  \bibinfo {author} {\bibfnamefont {M.~H.}\ \bibnamefont {de~Jong}}, \bibinfo
  {author} {\bibfnamefont {P.~G.}\ \bibnamefont {Steeneken}}, \bibinfo {author}
  {\bibfnamefont {M.~A.}\ \bibnamefont {Bessa}},\ and\ \bibinfo {author}
  {\bibfnamefont {R.~A.}\ \bibnamefont {Norte}},\ }\bibfield  {title} {\bibinfo
  {title} {Spiderweb nanomechanical resonators via bayesian optimization:
  inspired by nature and guided by machine learning},\ }\href@noop {}
  {\bibfield  {journal} {\bibinfo  {journal} {Advanced Materials}\ }\textbf
  {\bibinfo {volume} {34}},\ \bibinfo {pages} {2106248} (\bibinfo {year}
  {2022})}\BibitemShut {NoStop}%
\bibitem [{\citenamefont {Brandt}(1989)}]{brandt1989levitation}%
  \BibitemOpen
  \bibfield  {author} {\bibinfo {author} {\bibfnamefont {E.}~\bibnamefont
  {Brandt}},\ }\bibfield  {title} {\bibinfo {title} {Levitation in physics},\
  }\href@noop {} {\bibfield  {journal} {\bibinfo  {journal} {Science}\ }\textbf
  {\bibinfo {volume} {243}},\ \bibinfo {pages} {349} (\bibinfo {year}
  {1989})}\BibitemShut {NoStop}%
\bibitem [{\citenamefont {Gonzalez-Ballestero}\ \emph
  {et~al.}(2021)\citenamefont {Gonzalez-Ballestero}, \citenamefont
  {Aspelmeyer}, \citenamefont {Novotny}, \citenamefont {Quidant},\ and\
  \citenamefont {Romero-Isart}}]{gonzalez2021levitodynamics}%
  \BibitemOpen
  \bibfield  {author} {\bibinfo {author} {\bibfnamefont {C.}~\bibnamefont
  {Gonzalez-Ballestero}}, \bibinfo {author} {\bibfnamefont {M.}~\bibnamefont
  {Aspelmeyer}}, \bibinfo {author} {\bibfnamefont {L.}~\bibnamefont {Novotny}},
  \bibinfo {author} {\bibfnamefont {R.}~\bibnamefont {Quidant}},\ and\ \bibinfo
  {author} {\bibfnamefont {O.}~\bibnamefont {Romero-Isart}},\ }\bibfield
  {title} {\bibinfo {title} {Levitodynamics: Levitation and control of
  microscopic objects in vacuum},\ }\href@noop {} {\bibfield  {journal}
  {\bibinfo  {journal} {Science}\ }\textbf {\bibinfo {volume} {374}},\ \bibinfo
  {pages} {eabg3027} (\bibinfo {year} {2021})}\BibitemShut {NoStop}%
\bibitem [{\citenamefont {Gieseler}\ \emph {et~al.}(2012)\citenamefont
  {Gieseler}, \citenamefont {Deutsch}, \citenamefont {Quidant},\ and\
  \citenamefont {Novotny}}]{gieseler2012subkelvin}%
  \BibitemOpen
  \bibfield  {author} {\bibinfo {author} {\bibfnamefont {J.}~\bibnamefont
  {Gieseler}}, \bibinfo {author} {\bibfnamefont {B.}~\bibnamefont {Deutsch}},
  \bibinfo {author} {\bibfnamefont {R.}~\bibnamefont {Quidant}},\ and\ \bibinfo
  {author} {\bibfnamefont {L.}~\bibnamefont {Novotny}},\ }\bibfield  {title}
  {\bibinfo {title} {Subkelvin parametric feedback cooling of a laser-trapped
  nanoparticle},\ }\href@noop {} {\bibfield  {journal} {\bibinfo  {journal}
  {Physical review letters}\ }\textbf {\bibinfo {volume} {109}},\ \bibinfo
  {pages} {103603} (\bibinfo {year} {2012})}\BibitemShut {NoStop}%
\bibitem [{\citenamefont {Vovrosh}\ \emph {et~al.}(2017)\citenamefont
  {Vovrosh}, \citenamefont {Rashid}, \citenamefont {Hempston}, \citenamefont
  {Bateman}, \citenamefont {Paternostro},\ and\ \citenamefont
  {Ulbricht}}]{vovrosh2017parametric}%
  \BibitemOpen
  \bibfield  {author} {\bibinfo {author} {\bibfnamefont {J.}~\bibnamefont
  {Vovrosh}}, \bibinfo {author} {\bibfnamefont {M.}~\bibnamefont {Rashid}},
  \bibinfo {author} {\bibfnamefont {D.}~\bibnamefont {Hempston}}, \bibinfo
  {author} {\bibfnamefont {J.}~\bibnamefont {Bateman}}, \bibinfo {author}
  {\bibfnamefont {M.}~\bibnamefont {Paternostro}},\ and\ \bibinfo {author}
  {\bibfnamefont {H.}~\bibnamefont {Ulbricht}},\ }\bibfield  {title} {\bibinfo
  {title} {Parametric feedback cooling of levitated optomechanics in a
  parabolic mirror trap},\ }\href@noop {} {\bibfield  {journal} {\bibinfo
  {journal} {JOSA B}\ }\textbf {\bibinfo {volume} {34}},\ \bibinfo {pages}
  {1421} (\bibinfo {year} {2017})}\BibitemShut {NoStop}%
\bibitem [{\citenamefont {Vinante}\ \emph {et~al.}(2020)\citenamefont
  {Vinante}, \citenamefont {Falferi}, \citenamefont {Gasbarri}, \citenamefont
  {Setter}, \citenamefont {Timberlake},\ and\ \citenamefont
  {Ulbricht}}]{vinante2020ultralow}%
  \BibitemOpen
  \bibfield  {author} {\bibinfo {author} {\bibfnamefont {A.}~\bibnamefont
  {Vinante}}, \bibinfo {author} {\bibfnamefont {P.}~\bibnamefont {Falferi}},
  \bibinfo {author} {\bibfnamefont {G.}~\bibnamefont {Gasbarri}}, \bibinfo
  {author} {\bibfnamefont {A.}~\bibnamefont {Setter}}, \bibinfo {author}
  {\bibfnamefont {C.}~\bibnamefont {Timberlake}},\ and\ \bibinfo {author}
  {\bibfnamefont {H.}~\bibnamefont {Ulbricht}},\ }\bibfield  {title} {\bibinfo
  {title} {Ultralow mechanical damping with meissner-levitated ferromagnetic
  microparticles},\ }\href@noop {} {\bibfield  {journal} {\bibinfo  {journal}
  {Physical Review Applied}\ }\textbf {\bibinfo {volume} {13}},\ \bibinfo
  {pages} {064027} (\bibinfo {year} {2020})}\BibitemShut {NoStop}%
\bibitem [{\citenamefont {Pontin}\ \emph {et~al.}(2020)\citenamefont {Pontin},
  \citenamefont {Bullier}, \citenamefont {Toro{\v{s}}},\ and\ \citenamefont
  {Barker}}]{pontin2020ultranarrow}%
  \BibitemOpen
  \bibfield  {author} {\bibinfo {author} {\bibfnamefont {A.}~\bibnamefont
  {Pontin}}, \bibinfo {author} {\bibfnamefont {N.}~\bibnamefont {Bullier}},
  \bibinfo {author} {\bibfnamefont {M.}~\bibnamefont {Toro{\v{s}}}},\ and\
  \bibinfo {author} {\bibfnamefont {P.}~\bibnamefont {Barker}},\ }\bibfield
  {title} {\bibinfo {title} {Ultranarrow-linewidth levitated nano-oscillator
  for testing dissipative wave-function collapse},\ }\href@noop {} {\bibfield
  {journal} {\bibinfo  {journal} {Physical Review Research}\ }\textbf {\bibinfo
  {volume} {2}},\ \bibinfo {pages} {023349} (\bibinfo {year}
  {2020})}\BibitemShut {NoStop}%
\bibitem [{\citenamefont {Simon}\ and\ \citenamefont
  {Geim}(2000)}]{simon2000diamagnetic}%
  \BibitemOpen
  \bibfield  {author} {\bibinfo {author} {\bibfnamefont {M.}~\bibnamefont
  {Simon}}\ and\ \bibinfo {author} {\bibfnamefont {A.}~\bibnamefont {Geim}},\
  }\bibfield  {title} {\bibinfo {title} {Diamagnetic levitation: Flying frogs
  and floating magnets},\ }\href@noop {} {\bibfield  {journal} {\bibinfo
  {journal} {Journal of applied physics}\ }\textbf {\bibinfo {volume} {87}},\
  \bibinfo {pages} {6200} (\bibinfo {year} {2000})}\BibitemShut {NoStop}%
\bibitem [{\citenamefont {Mirica}\ \emph {et~al.}(2011)\citenamefont {Mirica},
  \citenamefont {Ilievski}, \citenamefont {Ellerbee}, \citenamefont
  {Shevkoplyas},\ and\ \citenamefont {Whitesides}}]{1335771}%
  \BibitemOpen
  \bibfield  {author} {\bibinfo {author} {\bibfnamefont {K.~A.}\ \bibnamefont
  {Mirica}}, \bibinfo {author} {\bibfnamefont {F.}~\bibnamefont {Ilievski}},
  \bibinfo {author} {\bibfnamefont {A.~K.}\ \bibnamefont {Ellerbee}}, \bibinfo
  {author} {\bibfnamefont {S.~S.}\ \bibnamefont {Shevkoplyas}},\ and\ \bibinfo
  {author} {\bibfnamefont {G.~M.}\ \bibnamefont {Whitesides}},\ }\bibfield
  {title} {\bibinfo {title} {Using magnetic levitation for three dimensional
  self-assembly},\ }\href@noop {} {\bibfield  {journal} {\bibinfo  {journal}
  {Advanced Materials}\ }\textbf {\bibinfo {volume} {23}},\ \bibinfo {pages}
  {4134} (\bibinfo {year} {2011})},\ \bibinfo {note} {1128}\BibitemShut
  {NoStop}%
\bibitem [{\citenamefont {Chen}\ \emph {et~al.}(2020)\citenamefont {Chen},
  \citenamefont {Ke{\c{s}}kekler}, \citenamefont {Alijani},\ and\ \citenamefont
  {Steeneken}}]{chen2020rigid}%
  \BibitemOpen
  \bibfield  {author} {\bibinfo {author} {\bibfnamefont {X.}~\bibnamefont
  {Chen}}, \bibinfo {author} {\bibfnamefont {A.}~\bibnamefont
  {Ke{\c{s}}kekler}}, \bibinfo {author} {\bibfnamefont {F.}~\bibnamefont
  {Alijani}},\ and\ \bibinfo {author} {\bibfnamefont {P.~G.}\ \bibnamefont
  {Steeneken}},\ }\bibfield  {title} {\bibinfo {title} {Rigid body dynamics of
  diamagnetically levitating graphite resonators},\ }\href@noop {} {\bibfield
  {journal} {\bibinfo  {journal} {Applied Physics Letters}\ }\textbf {\bibinfo
  {volume} {116}},\ \bibinfo {pages} {243505} (\bibinfo {year}
  {2020})}\BibitemShut {NoStop}%
\bibitem [{\citenamefont {Chen}\ \emph {et~al.}(2021)\citenamefont {Chen},
  \citenamefont {Kothari}, \citenamefont {Ke{\c{s}}kekler}, \citenamefont
  {Steeneken},\ and\ \citenamefont {Alijani}}]{chen2021diamagnetically}%
  \BibitemOpen
  \bibfield  {author} {\bibinfo {author} {\bibfnamefont {X.}~\bibnamefont
  {Chen}}, \bibinfo {author} {\bibfnamefont {N.}~\bibnamefont {Kothari}},
  \bibinfo {author} {\bibfnamefont {A.}~\bibnamefont {Ke{\c{s}}kekler}},
  \bibinfo {author} {\bibfnamefont {P.~G.}\ \bibnamefont {Steeneken}},\ and\
  \bibinfo {author} {\bibfnamefont {F.}~\bibnamefont {Alijani}},\ }\bibfield
  {title} {\bibinfo {title} {Diamagnetically levitating resonant weighing
  scale},\ }\href@noop {} {\bibfield  {journal} {\bibinfo  {journal} {Sensors
  and Actuators A: Physical}\ }\textbf {\bibinfo {volume} {330}},\ \bibinfo
  {pages} {112842} (\bibinfo {year} {2021})}\BibitemShut {NoStop}%
\bibitem [{\citenamefont {Monteiro}\ \emph {et~al.}(2017)\citenamefont
  {Monteiro}, \citenamefont {Ghosh}, \citenamefont {Fine},\ and\ \citenamefont
  {Moore}}]{monteiro2017optical}%
  \BibitemOpen
  \bibfield  {author} {\bibinfo {author} {\bibfnamefont {F.}~\bibnamefont
  {Monteiro}}, \bibinfo {author} {\bibfnamefont {S.}~\bibnamefont {Ghosh}},
  \bibinfo {author} {\bibfnamefont {A.~G.}\ \bibnamefont {Fine}},\ and\
  \bibinfo {author} {\bibfnamefont {D.~C.}\ \bibnamefont {Moore}},\ }\bibfield
  {title} {\bibinfo {title} {Optical levitation of 10-ng spheres with nano-g
  acceleration sensitivity},\ }\href@noop {} {\bibfield  {journal} {\bibinfo
  {journal} {Physical Review A}\ }\textbf {\bibinfo {volume} {96}},\ \bibinfo
  {pages} {063841} (\bibinfo {year} {2017})}\BibitemShut {NoStop}%
\bibitem [{\citenamefont {Moore}\ and\ \citenamefont
  {Geraci}(2021)}]{moore2021searching}%
  \BibitemOpen
  \bibfield  {author} {\bibinfo {author} {\bibfnamefont {D.~C.}\ \bibnamefont
  {Moore}}\ and\ \bibinfo {author} {\bibfnamefont {A.~A.}\ \bibnamefont
  {Geraci}},\ }\bibfield  {title} {\bibinfo {title} {Searching for new physics
  using optically levitated sensors},\ }\href@noop {} {\bibfield  {journal}
  {\bibinfo  {journal} {Quantum Science and Technology}\ }\textbf {\bibinfo
  {volume} {6}},\ \bibinfo {pages} {014008} (\bibinfo {year}
  {2021})}\BibitemShut {NoStop}%
\bibitem [{\citenamefont {Timberlake}\ \emph {et~al.}(2019)\citenamefont
  {Timberlake}, \citenamefont {Gasbarri}, \citenamefont {Vinante},
  \citenamefont {Setter},\ and\ \citenamefont
  {Ulbricht}}]{timberlake2019acceleration}%
  \BibitemOpen
  \bibfield  {author} {\bibinfo {author} {\bibfnamefont {C.}~\bibnamefont
  {Timberlake}}, \bibinfo {author} {\bibfnamefont {G.}~\bibnamefont
  {Gasbarri}}, \bibinfo {author} {\bibfnamefont {A.}~\bibnamefont {Vinante}},
  \bibinfo {author} {\bibfnamefont {A.}~\bibnamefont {Setter}},\ and\ \bibinfo
  {author} {\bibfnamefont {H.}~\bibnamefont {Ulbricht}},\ }\bibfield  {title}
  {\bibinfo {title} {Acceleration sensing with magnetically levitated
  oscillators above a superconductor},\ }\href@noop {} {\bibfield  {journal}
  {\bibinfo  {journal} {Applied Physics Letters}\ }\textbf {\bibinfo {volume}
  {115}},\ \bibinfo {pages} {224101} (\bibinfo {year} {2019})}\BibitemShut
  {NoStop}%
\bibitem [{\citenamefont {Middlemiss}\ \emph {et~al.}(2016)\citenamefont
  {Middlemiss}, \citenamefont {Samarelli}, \citenamefont {Paul}, \citenamefont
  {Hough}, \citenamefont {Rowan},\ and\ \citenamefont
  {Hammond}}]{middlemiss2016measurement}%
  \BibitemOpen
  \bibfield  {author} {\bibinfo {author} {\bibfnamefont {R.}~\bibnamefont
  {Middlemiss}}, \bibinfo {author} {\bibfnamefont {A.}~\bibnamefont
  {Samarelli}}, \bibinfo {author} {\bibfnamefont {D.}~\bibnamefont {Paul}},
  \bibinfo {author} {\bibfnamefont {J.}~\bibnamefont {Hough}}, \bibinfo
  {author} {\bibfnamefont {S.}~\bibnamefont {Rowan}},\ and\ \bibinfo {author}
  {\bibfnamefont {G.}~\bibnamefont {Hammond}},\ }\bibfield  {title} {\bibinfo
  {title} {Measurement of the earth tides with a mems gravimeter},\ }\href@noop
  {} {\bibfield  {journal} {\bibinfo  {journal} {Nature}\ }\textbf {\bibinfo
  {volume} {531}},\ \bibinfo {pages} {614} (\bibinfo {year}
  {2016})}\BibitemShut {NoStop}%
\bibitem [{\citenamefont {Schm{\"o}le}\ \emph {et~al.}(2016)\citenamefont
  {Schm{\"o}le}, \citenamefont {Dragosits}, \citenamefont {Hepach},\ and\
  \citenamefont {Aspelmeyer}}]{schmole2016micromechanical}%
  \BibitemOpen
  \bibfield  {author} {\bibinfo {author} {\bibfnamefont {J.}~\bibnamefont
  {Schm{\"o}le}}, \bibinfo {author} {\bibfnamefont {M.}~\bibnamefont
  {Dragosits}}, \bibinfo {author} {\bibfnamefont {H.}~\bibnamefont {Hepach}},\
  and\ \bibinfo {author} {\bibfnamefont {M.}~\bibnamefont {Aspelmeyer}},\
  }\bibfield  {title} {\bibinfo {title} {A micromechanical proof-of-principle
  experiment for measuring the gravitational force of milligram masses},\
  }\href@noop {} {\bibfield  {journal} {\bibinfo  {journal} {Classical and
  Quantum Gravity}\ }\textbf {\bibinfo {volume} {33}},\ \bibinfo {pages}
  {125031} (\bibinfo {year} {2016})}\BibitemShut {NoStop}%
\bibitem [{\citenamefont {Marletto}\ and\ \citenamefont
  {Vedral}(2017)}]{marletto2017gravitationally}%
  \BibitemOpen
  \bibfield  {author} {\bibinfo {author} {\bibfnamefont {C.}~\bibnamefont
  {Marletto}}\ and\ \bibinfo {author} {\bibfnamefont {V.}~\bibnamefont
  {Vedral}},\ }\bibfield  {title} {\bibinfo {title} {Gravitationally induced
  entanglement between two massive particles is sufficient evidence of quantum
  effects in gravity},\ }\href@noop {} {\bibfield  {journal} {\bibinfo
  {journal} {Physical review letters}\ }\textbf {\bibinfo {volume} {119}},\
  \bibinfo {pages} {240402} (\bibinfo {year} {2017})}\BibitemShut {NoStop}%
\bibitem [{\citenamefont {Belenchia}\ \emph {et~al.}(2018)\citenamefont
  {Belenchia}, \citenamefont {Wald}, \citenamefont {Giacomini}, \citenamefont
  {Castro-Ruiz}, \citenamefont {Brukner},\ and\ \citenamefont
  {Aspelmeyer}}]{belenchia2018quantum}%
  \BibitemOpen
  \bibfield  {author} {\bibinfo {author} {\bibfnamefont {A.}~\bibnamefont
  {Belenchia}}, \bibinfo {author} {\bibfnamefont {R.~M.}\ \bibnamefont {Wald}},
  \bibinfo {author} {\bibfnamefont {F.}~\bibnamefont {Giacomini}}, \bibinfo
  {author} {\bibfnamefont {E.}~\bibnamefont {Castro-Ruiz}}, \bibinfo {author}
  {\bibfnamefont {{\v{C}}.}~\bibnamefont {Brukner}},\ and\ \bibinfo {author}
  {\bibfnamefont {M.}~\bibnamefont {Aspelmeyer}},\ }\bibfield  {title}
  {\bibinfo {title} {Quantum superposition of massive objects and the
  quantization of gravity},\ }\href@noop {} {\bibfield  {journal} {\bibinfo
  {journal} {Physical Review D}\ }\textbf {\bibinfo {volume} {98}},\ \bibinfo
  {pages} {126009} (\bibinfo {year} {2018})}\BibitemShut {NoStop}%
\bibitem [{\citenamefont {Slezak}\ \emph {et~al.}(2018)\citenamefont {Slezak},
  \citenamefont {Lewandowski}, \citenamefont {Hsu},\ and\ \citenamefont
  {D’Urso}}]{slezak2018cooling}%
  \BibitemOpen
  \bibfield  {author} {\bibinfo {author} {\bibfnamefont {B.~R.}\ \bibnamefont
  {Slezak}}, \bibinfo {author} {\bibfnamefont {C.~W.}\ \bibnamefont
  {Lewandowski}}, \bibinfo {author} {\bibfnamefont {J.-F.}\ \bibnamefont
  {Hsu}},\ and\ \bibinfo {author} {\bibfnamefont {B.}~\bibnamefont
  {D’Urso}},\ }\bibfield  {title} {\bibinfo {title} {Cooling the motion of a
  silica microsphere in a magneto-gravitational trap in ultra-high vacuum},\
  }\href@noop {} {\bibfield  {journal} {\bibinfo  {journal} {New Journal of
  Physics}\ }\textbf {\bibinfo {volume} {20}},\ \bibinfo {pages} {063028}
  (\bibinfo {year} {2018})}\BibitemShut {NoStop}%
\bibitem [{\citenamefont {Lewandowski}\ \emph {et~al.}(2021)\citenamefont
  {Lewandowski}, \citenamefont {Knowles}, \citenamefont {Etienne},\ and\
  \citenamefont {D’Urso}}]{lewandowski2021high}%
  \BibitemOpen
  \bibfield  {author} {\bibinfo {author} {\bibfnamefont {C.~W.}\ \bibnamefont
  {Lewandowski}}, \bibinfo {author} {\bibfnamefont {T.~D.}\ \bibnamefont
  {Knowles}}, \bibinfo {author} {\bibfnamefont {Z.~B.}\ \bibnamefont
  {Etienne}},\ and\ \bibinfo {author} {\bibfnamefont {B.}~\bibnamefont
  {D’Urso}},\ }\bibfield  {title} {\bibinfo {title} {High-sensitivity
  accelerometry with a feedback-cooled magnetically levitated microsphere},\
  }\href@noop {} {\bibfield  {journal} {\bibinfo  {journal} {Physical Review
  Applied}\ }\textbf {\bibinfo {volume} {15}},\ \bibinfo {pages} {014050}
  (\bibinfo {year} {2021})}\BibitemShut {NoStop}%
\bibitem [{\citenamefont {Taghvaei}\ \emph {et~al.}(2009)\citenamefont
  {Taghvaei}, \citenamefont {Shokrollahi}, \citenamefont {Janghorban},\ and\
  \citenamefont {Abiri}}]{taghvaei2009eddy}%
  \BibitemOpen
  \bibfield  {author} {\bibinfo {author} {\bibfnamefont {A.}~\bibnamefont
  {Taghvaei}}, \bibinfo {author} {\bibfnamefont {H.}~\bibnamefont
  {Shokrollahi}}, \bibinfo {author} {\bibfnamefont {K.}~\bibnamefont
  {Janghorban}},\ and\ \bibinfo {author} {\bibfnamefont {H.}~\bibnamefont
  {Abiri}},\ }\bibfield  {title} {\bibinfo {title} {Eddy current and total
  power loss separation in the iron--phosphate--polyepoxy soft magnetic
  composites},\ }\href@noop {} {\bibfield  {journal} {\bibinfo  {journal}
  {Materials \& Design}\ }\textbf {\bibinfo {volume} {30}},\ \bibinfo {pages}
  {3989} (\bibinfo {year} {2009})}\BibitemShut {NoStop}%
\bibitem [{\citenamefont {Li}\ \emph {et~al.}(2011)\citenamefont {Li},
  \citenamefont {Kheifets},\ and\ \citenamefont {Raizen}}]{li2011millikelvin}%
  \BibitemOpen
  \bibfield  {author} {\bibinfo {author} {\bibfnamefont {T.}~\bibnamefont
  {Li}}, \bibinfo {author} {\bibfnamefont {S.}~\bibnamefont {Kheifets}},\ and\
  \bibinfo {author} {\bibfnamefont {M.~G.}\ \bibnamefont {Raizen}},\ }\bibfield
   {title} {\bibinfo {title} {Millikelvin cooling of an optically trapped
  microsphere in vacuum},\ }\href@noop {} {\bibfield  {journal} {\bibinfo
  {journal} {Nature Physics}\ }\textbf {\bibinfo {volume} {7}},\ \bibinfo
  {pages} {527} (\bibinfo {year} {2011})}\BibitemShut {NoStop}%
\bibitem [{\citenamefont {Ranjit}\ \emph {et~al.}(2016)\citenamefont {Ranjit},
  \citenamefont {Cunningham}, \citenamefont {Casey},\ and\ \citenamefont
  {Geraci}}]{ranjit2016zeptonewton}%
  \BibitemOpen
  \bibfield  {author} {\bibinfo {author} {\bibfnamefont {G.}~\bibnamefont
  {Ranjit}}, \bibinfo {author} {\bibfnamefont {M.}~\bibnamefont {Cunningham}},
  \bibinfo {author} {\bibfnamefont {K.}~\bibnamefont {Casey}},\ and\ \bibinfo
  {author} {\bibfnamefont {A.~A.}\ \bibnamefont {Geraci}},\ }\bibfield  {title}
  {\bibinfo {title} {Zeptonewton force sensing with nanospheres in an optical
  lattice},\ }\href@noop {} {\bibfield  {journal} {\bibinfo  {journal}
  {Physical Review A}\ }\textbf {\bibinfo {volume} {93}},\ \bibinfo {pages}
  {053801} (\bibinfo {year} {2016})}\BibitemShut {NoStop}%
\bibitem [{\citenamefont {Ranjit}\ \emph {et~al.}(2015)\citenamefont {Ranjit},
  \citenamefont {Atherton}, \citenamefont {Stutz}, \citenamefont {Cunningham},\
  and\ \citenamefont {Geraci}}]{ranjit2015attonewton}%
  \BibitemOpen
  \bibfield  {author} {\bibinfo {author} {\bibfnamefont {G.}~\bibnamefont
  {Ranjit}}, \bibinfo {author} {\bibfnamefont {D.~P.}\ \bibnamefont
  {Atherton}}, \bibinfo {author} {\bibfnamefont {J.~H.}\ \bibnamefont {Stutz}},
  \bibinfo {author} {\bibfnamefont {M.}~\bibnamefont {Cunningham}},\ and\
  \bibinfo {author} {\bibfnamefont {A.~A.}\ \bibnamefont {Geraci}},\ }\bibfield
   {title} {\bibinfo {title} {Attonewton force detection using microspheres in
  a dual-beam optical trap in high vacuum},\ }\href@noop {} {\bibfield
  {journal} {\bibinfo  {journal} {Physical Review A}\ }\textbf {\bibinfo
  {volume} {91}},\ \bibinfo {pages} {051805} (\bibinfo {year}
  {2015})}\BibitemShut {NoStop}%
\bibitem [{\citenamefont {Leng}\ \emph {et~al.}(2021)\citenamefont {Leng},
  \citenamefont {Li}, \citenamefont {Kong}, \citenamefont {Xie}, \citenamefont
  {Zheng}, \citenamefont {Yin}, \citenamefont {Xiong}, \citenamefont {Wu},
  \citenamefont {Duan}, \citenamefont {Du} \emph
  {et~al.}}]{leng2021mechanical}%
  \BibitemOpen
  \bibfield  {author} {\bibinfo {author} {\bibfnamefont {Y.}~\bibnamefont
  {Leng}}, \bibinfo {author} {\bibfnamefont {R.}~\bibnamefont {Li}}, \bibinfo
  {author} {\bibfnamefont {X.}~\bibnamefont {Kong}}, \bibinfo {author}
  {\bibfnamefont {H.}~\bibnamefont {Xie}}, \bibinfo {author} {\bibfnamefont
  {D.}~\bibnamefont {Zheng}}, \bibinfo {author} {\bibfnamefont
  {P.}~\bibnamefont {Yin}}, \bibinfo {author} {\bibfnamefont {F.}~\bibnamefont
  {Xiong}}, \bibinfo {author} {\bibfnamefont {T.}~\bibnamefont {Wu}}, \bibinfo
  {author} {\bibfnamefont {C.-K.}\ \bibnamefont {Duan}}, \bibinfo {author}
  {\bibfnamefont {Y.}~\bibnamefont {Du}}, \emph {et~al.},\ }\bibfield  {title}
  {\bibinfo {title} {Mechanical dissipation below 1 $\mu$ hz with a cryogenic
  diamagnetic levitated micro-oscillator},\ }\href@noop {} {\bibfield
  {journal} {\bibinfo  {journal} {Physical Review Applied}\ }\textbf {\bibinfo
  {volume} {15}},\ \bibinfo {pages} {024061} (\bibinfo {year}
  {2021})}\BibitemShut {NoStop}%
\bibitem [{\citenamefont {Xiong}\ \emph {et~al.}(2021)\citenamefont {Xiong},
  \citenamefont {Yin}, \citenamefont {Wu}, \citenamefont {Xie}, \citenamefont
  {Li}, \citenamefont {Leng}, \citenamefont {Li}, \citenamefont {Duan},
  \citenamefont {Kong}, \citenamefont {Huang},\ and\ \citenamefont
  {Du}}]{PhysRevApplied.16.L011003}%
  \BibitemOpen
  \bibfield  {author} {\bibinfo {author} {\bibfnamefont {F.}~\bibnamefont
  {Xiong}}, \bibinfo {author} {\bibfnamefont {P.}~\bibnamefont {Yin}}, \bibinfo
  {author} {\bibfnamefont {T.}~\bibnamefont {Wu}}, \bibinfo {author}
  {\bibfnamefont {H.}~\bibnamefont {Xie}}, \bibinfo {author} {\bibfnamefont
  {R.}~\bibnamefont {Li}}, \bibinfo {author} {\bibfnamefont {Y.}~\bibnamefont
  {Leng}}, \bibinfo {author} {\bibfnamefont {Y.}~\bibnamefont {Li}}, \bibinfo
  {author} {\bibfnamefont {C.}~\bibnamefont {Duan}}, \bibinfo {author}
  {\bibfnamefont {X.}~\bibnamefont {Kong}}, \bibinfo {author} {\bibfnamefont
  {P.}~\bibnamefont {Huang}},\ and\ \bibinfo {author} {\bibfnamefont
  {J.}~\bibnamefont {Du}},\ }\bibfield  {title} {\bibinfo {title} {Lens-free
  optical detection of thermal motion of a submillimeter sphere diamagnetically
  levitated in high vacuum},\ }\href
  {https://doi.org/10.1103/PhysRevApplied.16.L011003} {\bibfield  {journal}
  {\bibinfo  {journal} {Phys. Rev. Applied}\ }\textbf {\bibinfo {volume}
  {16}},\ \bibinfo {pages} {L011003} (\bibinfo {year} {2021})}\BibitemShut
  {NoStop}%
\bibitem [{\citenamefont {Gieseler}\ \emph {et~al.}(2020)\citenamefont
  {Gieseler}, \citenamefont {Kabcenell}, \citenamefont {Rosenfeld},
  \citenamefont {Schaefer}, \citenamefont {Safira}, \citenamefont {Schuetz},
  \citenamefont {Gonzalez-Ballestero}, \citenamefont {Rusconi}, \citenamefont
  {Romero-Isart},\ and\ \citenamefont {Lukin}}]{gieseler2020single}%
  \BibitemOpen
  \bibfield  {author} {\bibinfo {author} {\bibfnamefont {J.}~\bibnamefont
  {Gieseler}}, \bibinfo {author} {\bibfnamefont {A.}~\bibnamefont {Kabcenell}},
  \bibinfo {author} {\bibfnamefont {E.}~\bibnamefont {Rosenfeld}}, \bibinfo
  {author} {\bibfnamefont {J.}~\bibnamefont {Schaefer}}, \bibinfo {author}
  {\bibfnamefont {A.}~\bibnamefont {Safira}}, \bibinfo {author} {\bibfnamefont
  {M.~J.}\ \bibnamefont {Schuetz}}, \bibinfo {author} {\bibfnamefont
  {C.}~\bibnamefont {Gonzalez-Ballestero}}, \bibinfo {author} {\bibfnamefont
  {C.~C.}\ \bibnamefont {Rusconi}}, \bibinfo {author} {\bibfnamefont
  {O.}~\bibnamefont {Romero-Isart}},\ and\ \bibinfo {author} {\bibfnamefont
  {M.~D.}\ \bibnamefont {Lukin}},\ }\bibfield  {title} {\bibinfo {title}
  {Single-spin magnetomechanics with levitated micromagnets},\ }\href@noop {}
  {\bibfield  {journal} {\bibinfo  {journal} {Physical Review Letters}\
  }\textbf {\bibinfo {volume} {124}},\ \bibinfo {pages} {163604} (\bibinfo
  {year} {2020})}\BibitemShut {NoStop}%
\bibitem [{\citenamefont {Wang}\ \emph {et~al.}(2019)\citenamefont {Wang},
  \citenamefont {Lourette}, \citenamefont {O’Kelley}, \citenamefont {Kayci},
  \citenamefont {Band}, \citenamefont {Kimball}, \citenamefont {Sushkov},\ and\
  \citenamefont {Budker}}]{wang2019dynamics}%
  \BibitemOpen
  \bibfield  {author} {\bibinfo {author} {\bibfnamefont {T.}~\bibnamefont
  {Wang}}, \bibinfo {author} {\bibfnamefont {S.}~\bibnamefont {Lourette}},
  \bibinfo {author} {\bibfnamefont {S.~R.}\ \bibnamefont {O’Kelley}},
  \bibinfo {author} {\bibfnamefont {M.}~\bibnamefont {Kayci}}, \bibinfo
  {author} {\bibfnamefont {Y.}~\bibnamefont {Band}}, \bibinfo {author}
  {\bibfnamefont {D.~F.~J.}\ \bibnamefont {Kimball}}, \bibinfo {author}
  {\bibfnamefont {A.~O.}\ \bibnamefont {Sushkov}},\ and\ \bibinfo {author}
  {\bibfnamefont {D.}~\bibnamefont {Budker}},\ }\bibfield  {title} {\bibinfo
  {title} {Dynamics of a ferromagnetic particle levitated over a
  superconductor},\ }\href@noop {} {\bibfield  {journal} {\bibinfo  {journal}
  {Physical Review Applied}\ }\textbf {\bibinfo {volume} {11}},\ \bibinfo
  {pages} {044041} (\bibinfo {year} {2019})}\BibitemShut {NoStop}%
\bibitem [{\citenamefont {Hempston}\ \emph {et~al.}(2017)\citenamefont
  {Hempston}, \citenamefont {Vovrosh}, \citenamefont {Toro{\v{s}}},
  \citenamefont {Winstone}, \citenamefont {Rashid},\ and\ \citenamefont
  {Ulbricht}}]{hempston2017force}%
  \BibitemOpen
  \bibfield  {author} {\bibinfo {author} {\bibfnamefont {D.}~\bibnamefont
  {Hempston}}, \bibinfo {author} {\bibfnamefont {J.}~\bibnamefont {Vovrosh}},
  \bibinfo {author} {\bibfnamefont {M.}~\bibnamefont {Toro{\v{s}}}}, \bibinfo
  {author} {\bibfnamefont {G.}~\bibnamefont {Winstone}}, \bibinfo {author}
  {\bibfnamefont {M.}~\bibnamefont {Rashid}},\ and\ \bibinfo {author}
  {\bibfnamefont {H.}~\bibnamefont {Ulbricht}},\ }\bibfield  {title} {\bibinfo
  {title} {Force sensing with an optically levitated charged nanoparticle},\
  }\href@noop {} {\bibfield  {journal} {\bibinfo  {journal} {Applied Physics
  Letters}\ }\textbf {\bibinfo {volume} {111}},\ \bibinfo {pages} {133111}
  (\bibinfo {year} {2017})}\BibitemShut {NoStop}%
\bibitem [{\citenamefont {Kawasaki}\ \emph {et~al.}(2020)\citenamefont
  {Kawasaki}, \citenamefont {Fieguth}, \citenamefont {Priel}, \citenamefont
  {Blakemore}, \citenamefont {Martin},\ and\ \citenamefont
  {Gratta}}]{kawasaki2020high}%
  \BibitemOpen
  \bibfield  {author} {\bibinfo {author} {\bibfnamefont {A.}~\bibnamefont
  {Kawasaki}}, \bibinfo {author} {\bibfnamefont {A.}~\bibnamefont {Fieguth}},
  \bibinfo {author} {\bibfnamefont {N.}~\bibnamefont {Priel}}, \bibinfo
  {author} {\bibfnamefont {C.~P.}\ \bibnamefont {Blakemore}}, \bibinfo {author}
  {\bibfnamefont {D.}~\bibnamefont {Martin}},\ and\ \bibinfo {author}
  {\bibfnamefont {G.}~\bibnamefont {Gratta}},\ }\bibfield  {title} {\bibinfo
  {title} {High sensitivity, levitated microsphere apparatus for short-distance
  force measurements},\ }\href@noop {} {\bibfield  {journal} {\bibinfo
  {journal} {Review of Scientific Instruments}\ }\textbf {\bibinfo {volume}
  {91}},\ \bibinfo {pages} {083201} (\bibinfo {year} {2020})}\BibitemShut
  {NoStop}%
\bibitem [{\citenamefont {Moore}\ \emph {et~al.}(2014)\citenamefont {Moore},
  \citenamefont {Rider},\ and\ \citenamefont {Gratta}}]{moore2014search}%
  \BibitemOpen
  \bibfield  {author} {\bibinfo {author} {\bibfnamefont {D.~C.}\ \bibnamefont
  {Moore}}, \bibinfo {author} {\bibfnamefont {A.~D.}\ \bibnamefont {Rider}},\
  and\ \bibinfo {author} {\bibfnamefont {G.}~\bibnamefont {Gratta}},\
  }\bibfield  {title} {\bibinfo {title} {Search for millicharged particles
  using optically levitated microspheres},\ }\href@noop {} {\bibfield
  {journal} {\bibinfo  {journal} {Physical review letters}\ }\textbf {\bibinfo
  {volume} {113}},\ \bibinfo {pages} {251801} (\bibinfo {year}
  {2014})}\BibitemShut {NoStop}%
\bibitem [{\citenamefont {Rider}\ \emph {et~al.}(2018)\citenamefont {Rider},
  \citenamefont {Blakemore}, \citenamefont {Gratta},\ and\ \citenamefont
  {Moore}}]{rider2018single}%
  \BibitemOpen
  \bibfield  {author} {\bibinfo {author} {\bibfnamefont {A.~D.}\ \bibnamefont
  {Rider}}, \bibinfo {author} {\bibfnamefont {C.~P.}\ \bibnamefont
  {Blakemore}}, \bibinfo {author} {\bibfnamefont {G.}~\bibnamefont {Gratta}},\
  and\ \bibinfo {author} {\bibfnamefont {D.~C.}\ \bibnamefont {Moore}},\
  }\bibfield  {title} {\bibinfo {title} {Single-beam dielectric-microsphere
  trapping with optical heterodyne detection},\ }\href@noop {} {\bibfield
  {journal} {\bibinfo  {journal} {Physical Review A}\ }\textbf {\bibinfo
  {volume} {97}},\ \bibinfo {pages} {013842} (\bibinfo {year}
  {2018})}\BibitemShut {NoStop}%
\bibitem [{\citenamefont {Monteiro}\ \emph {et~al.}(2020)\citenamefont
  {Monteiro}, \citenamefont {Li}, \citenamefont {Afek}, \citenamefont {Li},
  \citenamefont {Mossman},\ and\ \citenamefont {Moore}}]{monteiro2020force}%
  \BibitemOpen
  \bibfield  {author} {\bibinfo {author} {\bibfnamefont {F.}~\bibnamefont
  {Monteiro}}, \bibinfo {author} {\bibfnamefont {W.}~\bibnamefont {Li}},
  \bibinfo {author} {\bibfnamefont {G.}~\bibnamefont {Afek}}, \bibinfo {author}
  {\bibfnamefont {C.-l.}\ \bibnamefont {Li}}, \bibinfo {author} {\bibfnamefont
  {M.}~\bibnamefont {Mossman}},\ and\ \bibinfo {author} {\bibfnamefont {D.~C.}\
  \bibnamefont {Moore}},\ }\bibfield  {title} {\bibinfo {title} {Force and
  acceleration sensing with optically levitated nanogram masses at microkelvin
  temperatures},\ }\href@noop {} {\bibfield  {journal} {\bibinfo  {journal}
  {Physical Review A}\ }\textbf {\bibinfo {volume} {101}},\ \bibinfo {pages}
  {053835} (\bibinfo {year} {2020})}\BibitemShut {NoStop}%
\bibitem [{\citenamefont {Whittle}\ \emph {et~al.}(2021)\citenamefont
  {Whittle}, \citenamefont {Hall}, \citenamefont {Dwyer}, \citenamefont
  {Mavalvala}, \citenamefont {Sudhir}, \citenamefont {Abbott}, \citenamefont
  {Ananyeva}, \citenamefont {Austin}, \citenamefont {Barsotti}, \citenamefont
  {Betzwieser} \emph {et~al.}}]{whittle2021approaching}%
  \BibitemOpen
  \bibfield  {author} {\bibinfo {author} {\bibfnamefont {C.}~\bibnamefont
  {Whittle}}, \bibinfo {author} {\bibfnamefont {E.~D.}\ \bibnamefont {Hall}},
  \bibinfo {author} {\bibfnamefont {S.}~\bibnamefont {Dwyer}}, \bibinfo
  {author} {\bibfnamefont {N.}~\bibnamefont {Mavalvala}}, \bibinfo {author}
  {\bibfnamefont {V.}~\bibnamefont {Sudhir}}, \bibinfo {author} {\bibfnamefont
  {R.}~\bibnamefont {Abbott}}, \bibinfo {author} {\bibfnamefont
  {A.}~\bibnamefont {Ananyeva}}, \bibinfo {author} {\bibfnamefont
  {C.}~\bibnamefont {Austin}}, \bibinfo {author} {\bibfnamefont
  {L.}~\bibnamefont {Barsotti}}, \bibinfo {author} {\bibfnamefont
  {J.}~\bibnamefont {Betzwieser}}, \emph {et~al.},\ }\bibfield  {title}
  {\bibinfo {title} {Approaching the motional ground state of a 10-kg object},\
  }\href@noop {} {\bibfield  {journal} {\bibinfo  {journal} {Science}\ }\textbf
  {\bibinfo {volume} {372}},\ \bibinfo {pages} {1333} (\bibinfo {year}
  {2021})}\BibitemShut {NoStop}%
\bibitem [{\citenamefont {Michimura}\ and\ \citenamefont
  {Komori}(2020)}]{michimura2020quantum}%
  \BibitemOpen
  \bibfield  {author} {\bibinfo {author} {\bibfnamefont {Y.}~\bibnamefont
  {Michimura}}\ and\ \bibinfo {author} {\bibfnamefont {K.}~\bibnamefont
  {Komori}},\ }\bibfield  {title} {\bibinfo {title} {Quantum sensing with
  milligram scale optomechanical systems},\ }\href@noop {} {\bibfield
  {journal} {\bibinfo  {journal} {The European Physical Journal D}\ }\textbf
  {\bibinfo {volume} {74}},\ \bibinfo {pages} {1} (\bibinfo {year}
  {2020})}\BibitemShut {NoStop}%
\bibitem [{\citenamefont {Yu}\ \emph {et~al.}(2020)\citenamefont {Yu},
  \citenamefont {McCuller}, \citenamefont {Tse}, \citenamefont {Kijbunchoo},
  \citenamefont {Barsotti},\ and\ \citenamefont {Mavalvala}}]{yu2020quantum}%
  \BibitemOpen
  \bibfield  {author} {\bibinfo {author} {\bibfnamefont {H.}~\bibnamefont
  {Yu}}, \bibinfo {author} {\bibfnamefont {L.}~\bibnamefont {McCuller}},
  \bibinfo {author} {\bibfnamefont {M.}~\bibnamefont {Tse}}, \bibinfo {author}
  {\bibfnamefont {N.}~\bibnamefont {Kijbunchoo}}, \bibinfo {author}
  {\bibfnamefont {L.}~\bibnamefont {Barsotti}},\ and\ \bibinfo {author}
  {\bibfnamefont {N.}~\bibnamefont {Mavalvala}},\ }\bibfield  {title} {\bibinfo
  {title} {Quantum correlations between light and the kilogram-mass mirrors of
  ligo},\ }\href@noop {} {\bibfield  {journal} {\bibinfo  {journal} {Nature}\
  }\textbf {\bibinfo {volume} {583}},\ \bibinfo {pages} {43} (\bibinfo {year}
  {2020})}\BibitemShut {NoStop}%
\bibitem [{\citenamefont {Bray}\ \emph {et~al.}(2011)\citenamefont {Bray},
  \citenamefont {Gilmour}, \citenamefont {Guild}, \citenamefont {Hsieh},
  \citenamefont {Masania},\ and\ \citenamefont {Taylor}}]{bray2011quantifying}%
  \BibitemOpen
  \bibfield  {author} {\bibinfo {author} {\bibfnamefont {D.}~\bibnamefont
  {Bray}}, \bibinfo {author} {\bibfnamefont {S.}~\bibnamefont {Gilmour}},
  \bibinfo {author} {\bibfnamefont {F.}~\bibnamefont {Guild}}, \bibinfo
  {author} {\bibfnamefont {T.}~\bibnamefont {Hsieh}}, \bibinfo {author}
  {\bibfnamefont {K.}~\bibnamefont {Masania}},\ and\ \bibinfo {author}
  {\bibfnamefont {A.}~\bibnamefont {Taylor}},\ }\bibfield  {title} {\bibinfo
  {title} {Quantifying nanoparticle dispersion: application of the delaunay
  network for objective analysis of sample micrographs},\ }\href@noop {}
  {\bibfield  {journal} {\bibinfo  {journal} {Journal of materials science}\
  }\textbf {\bibinfo {volume} {46}},\ \bibinfo {pages} {6437} (\bibinfo {year}
  {2011})}\BibitemShut {NoStop}%
\bibitem [{\citenamefont {Pappis}\ and\ \citenamefont
  {Blum}(1961)}]{pappis1961properties}%
  \BibitemOpen
  \bibfield  {author} {\bibinfo {author} {\bibfnamefont {J.}~\bibnamefont
  {Pappis}}\ and\ \bibinfo {author} {\bibfnamefont {S.}~\bibnamefont {Blum}},\
  }\bibfield  {title} {\bibinfo {title} {Properties of pyrolytic graphite},\
  }\href@noop {} {\bibfield  {journal} {\bibinfo  {journal} {Journal of the
  American Ceramic Society}\ }\textbf {\bibinfo {volume} {44}},\ \bibinfo
  {pages} {592} (\bibinfo {year} {1961})}\BibitemShut {NoStop}%
\bibitem [{\citenamefont {Boukallel}\ \emph {et~al.}(2003)\citenamefont
  {Boukallel}, \citenamefont {Abadie},\ and\ \citenamefont
  {Piat}}]{boukallel2003levitated}%
  \BibitemOpen
  \bibfield  {author} {\bibinfo {author} {\bibfnamefont {M.}~\bibnamefont
  {Boukallel}}, \bibinfo {author} {\bibfnamefont {J.}~\bibnamefont {Abadie}},\
  and\ \bibinfo {author} {\bibfnamefont {E.}~\bibnamefont {Piat}},\ }\bibfield
  {title} {\bibinfo {title} {Levitated micro-nano force sensor using
  diamagnetic materials},\ }in\ \href@noop {} {\emph {\bibinfo {booktitle}
  {2003 IEEE International Conference on Robotics and Automation (Cat. No.
  03CH37422)}}},\ Vol.~\bibinfo {volume} {3}\ (\bibinfo {organization} {IEEE},\
  \bibinfo {year} {2003})\ pp.\ \bibinfo {pages} {3219--3224}\BibitemShut
  {NoStop}%
\end{thebibliography}%

\onecolumngrid
\vspace{1cm}
\begin{center}

\large \textbf{Supplementary Information}
    
\end{center}
\vspace{0.5cm}

In S1, we show the details of graphite composite fabrication process. S2 details out particle size measurement (S2.1) and particle dispersion analysis (S2.2). S3 presents the FEM methodology used to calculate the eddy current damping in a graphite (S3.1) and a composite plate (S3.2). S4 shows the electrical conductivity measurements of composites with different volume fractions. In S5, we present the analytical model to calculate the eddy current damping of a composite plate moving in a magnetic field. Finally, in S6 we show the $Q$s and acceleration noise floor of state-of-the-art levitodynamic systems. 
%\tableofcontents
%\newpage
\section*{S1: Composite fabrication process}
\setcounter{figure}{0}
\begin{figure}[h!]
    \centering
    \renewcommand{\thefigure}{S\arabic{figure}}
    \includegraphics[height=3.5cm]{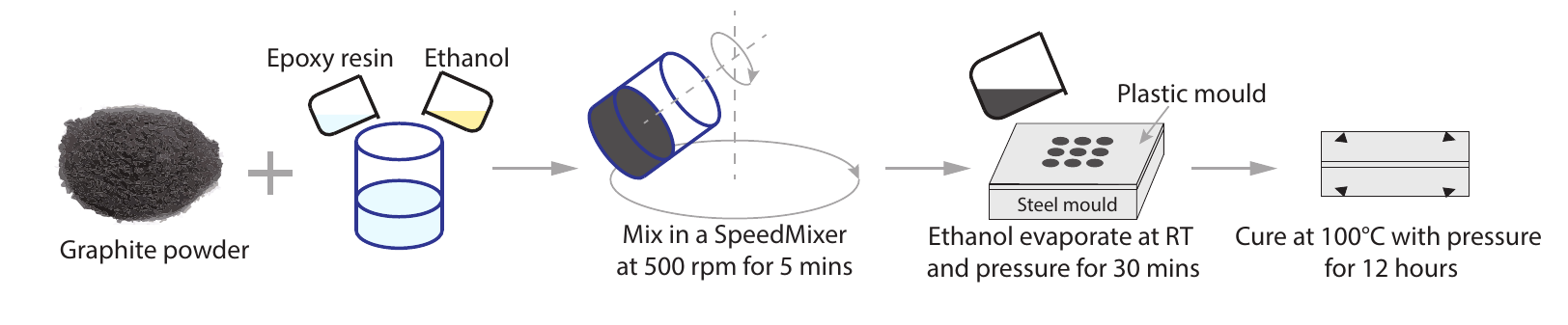}
    \caption{\label{fig:fabrication}
    The schematic of composite fabrication process.}
\end{figure}
The details of graphite composite fabrication process is shown in Fig. \ref{fig:fabrication}. First, the two components of the epoxy are mixed at 3500 rpm for 5 minutes in a  Dual Asymmetric Centrifuge mixer (DAC 150.1 FVZ-K ) followed by the addition and mixing of the  graphite powder at 500 rpm for 5 minutes. To reduce the viscosity of the resulting graphite-epoxy paste, ethanol is added, and further mixed at 500 rpm for \SI{5}{mins}. This maximises dispersion and homogeneity of the paste with the graphite particles in the epoxy-ethanol matrix. The paste is then transferred into circular holes ($\phi=\SI{10}{mm}$) in a thin plastic mould with thickness of \SI{0.12}{mm} on the top of a flat steel mould. The deposited paste is left at room temperature and pressure for 30 minutes to let the ethanol fully evaporate before curing the epoxy to minimise porosity. Afterwards, the graphite/epoxy paste is compressed by steel moulds with pressure, which is then cured in an oven at \SI{100}{\degree C} for around \SI{12}{hours}. Once the graphite/epoxy composite is cured, we use a micro laser cutter to cut the composite into square plates with desired lengths and use fine sand paper (\SI{5}{\um} grain) to polish its surface to the desired thickness.

%\newpage
\section*{S2: Particle size measurement and particle dispersion analysis}

\subsection*{S2.1: Particle size measurement}
\begin{figure}[h!]
    \centering
    \renewcommand{\thefigure}{S\arabic{figure}}
    \includegraphics[height=6cm]{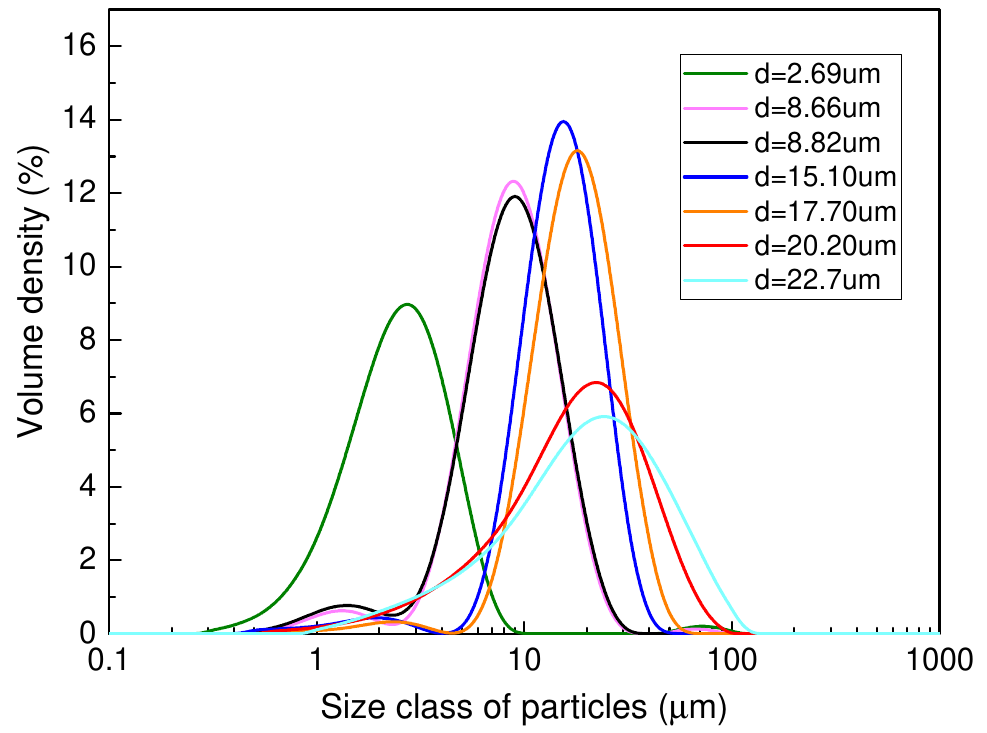}
    \caption{\label{fig:particle size}
    Particle size distribution for different particles used in the experiments.}
\end{figure}
Fig. \ref{fig:particle size} shows the size distribution of the graphite powders used in our experiments. The particle size distribution measurements are performed using Malvern Mastersizer 3000 on {0.1}\% w/v aqueous solution of the powders using sodium dodecyl sulfate solution as surfactant. The measurement of each type of powders is repeated five times. From this figure, it can be seen that the particle size of each type of powder has a wide range of distribution. In the main text, we use the mean value of the distribution to represent the particle size $d$.

\subsection*{S2.2:Particle dispersion analysis}
\begin{figure}[h!]
    \centering
    \renewcommand{\thefigure}{S\arabic{figure}}
    \includegraphics[height=8cm]{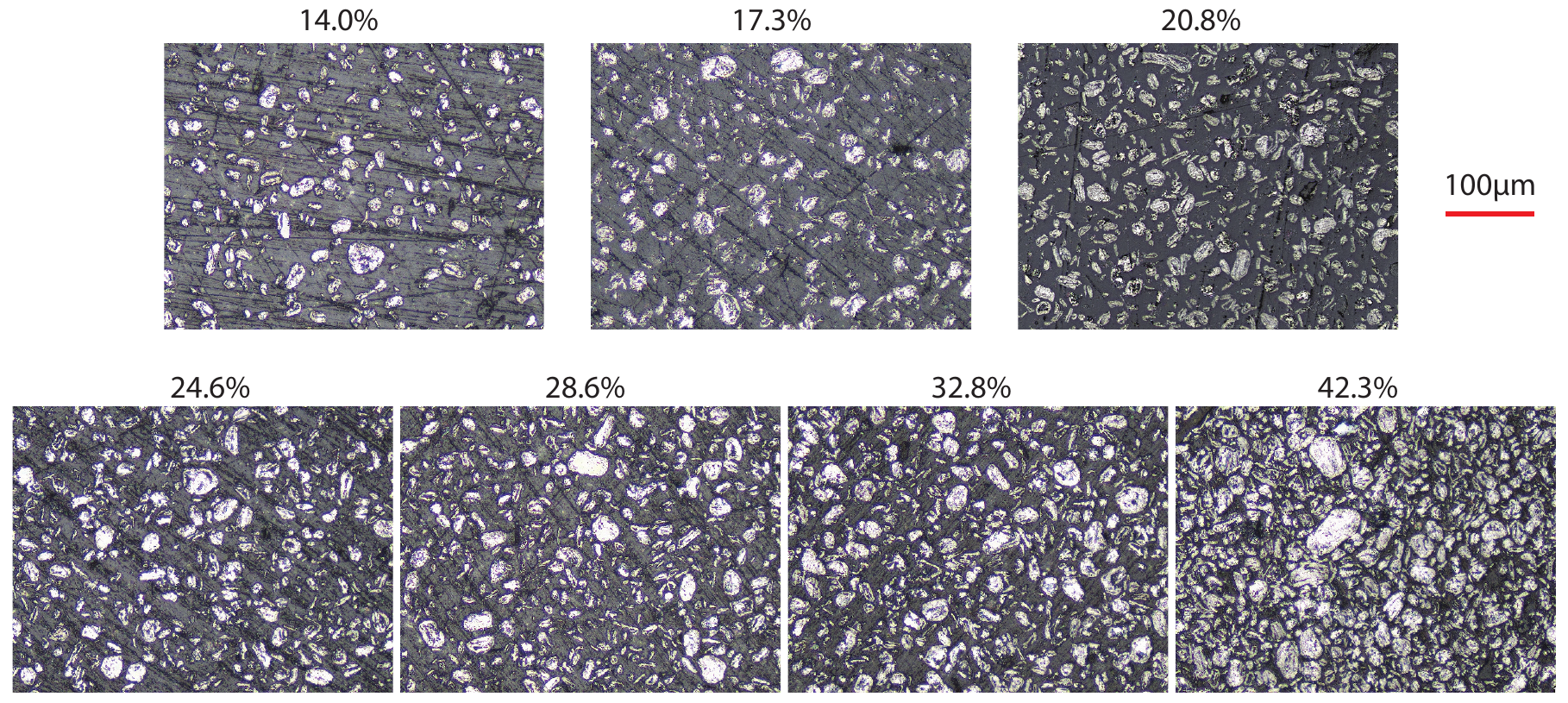}
    \caption{\label{fig:microscopy images}
    Microscopic images of composites with \SI{17.6}{\um} particles and different volume fractions.}
\end{figure}
\begin{figure}[h!]
    \centering
    \renewcommand{\thefigure}{S\arabic{figure}}
    \includegraphics[height=5cm]{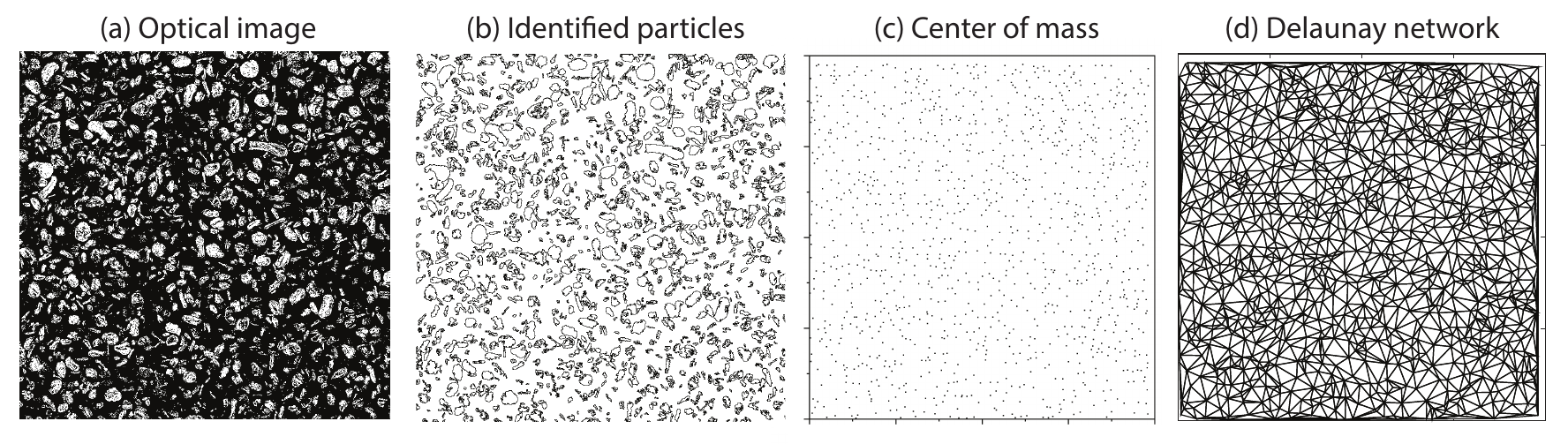}
    \caption{\label{fig:dispersion}
     Particle dispersion analysis on a composite with \SI{17.6}{\um} particles and \SI{20.8}{\percent} volume fraction. To obtain the Area Disorder $AD$, we first take an image of the composite (a) and then identify the particles (b) using ImageJ. Afterwards, the particles' center of mass (c) is located, from which we build their Delaunay network (d) to calculate the $AD$.}
\end{figure}
\begin{figure}[h!]
    \centering
    \renewcommand{\thefigure}{S\arabic{figure}}
    \includegraphics[height=8cm]{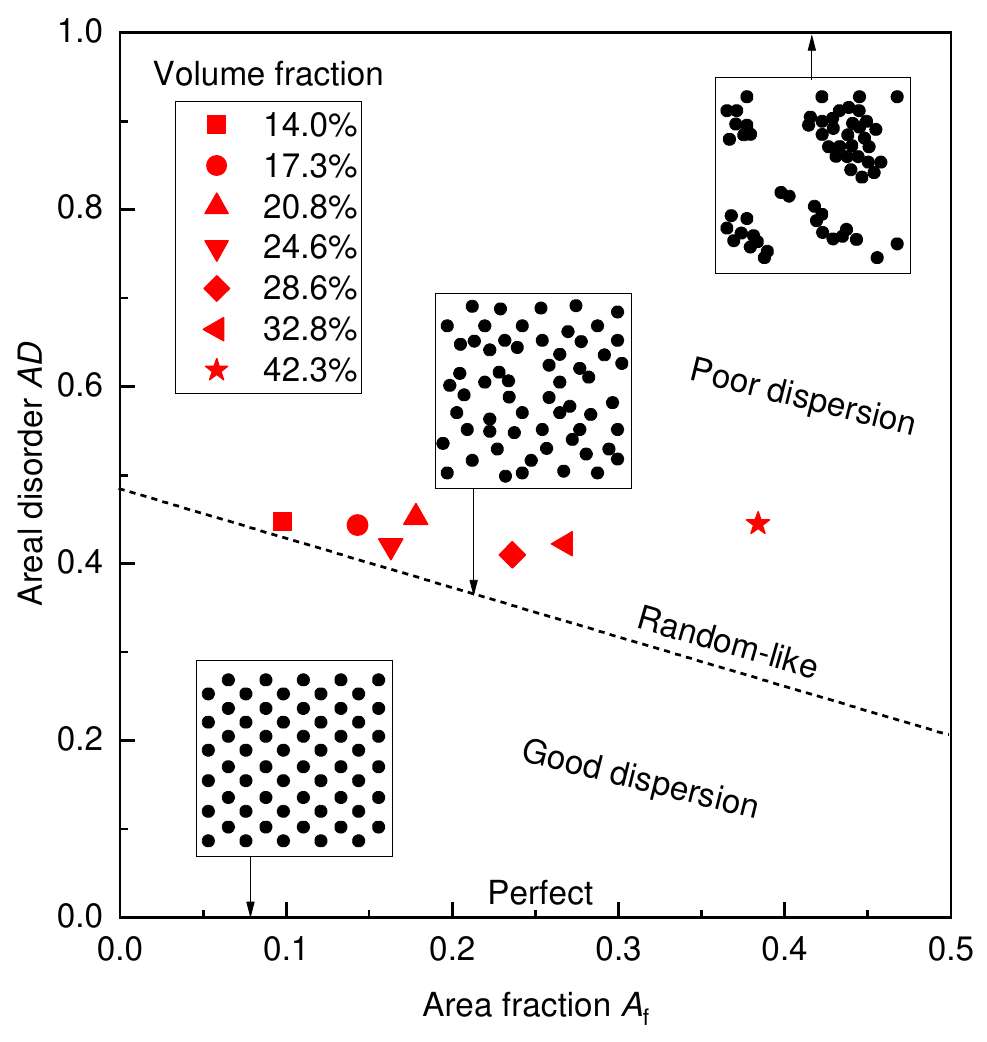}
    \caption{\label{fig:areal disorder}
    Quantitative analysis of the particle dispersion in composites with different volume fractions. The Area fraction $A_f$ is obtained via image processing.}
\end{figure}
Fig. \ref{fig:microscopy images} shows microscopic images of composites with \SI{17.6}{\um} particles and different volume fractions. To obtain these images, samples are first polished with a fine sand paper to obtain a clear interface between the graphite particles and epoxy. The graphite particles are appearing in white color due to the reflection of light from the microscope. To quantify the dispersion quality of the particles inside epoxy, we use the Area Disorder ($AD$) of the Delaunay network as described in reference \cite{bray2011quantifying}. $AD$ is a dimensionless quantity with values between 0 and 1. $AD=0$ means the dispersion is perfect and particles are homogeneously distributed inside the matrix. $AD=1$ means the dispersion is worst with clusters, as shown in Fig. \ref{fig:areal disorder}. To obtain the $AD$, we first identify the particle boundaries (Fig. \ref{fig:dispersion}b) from the optical image (Fig. \ref{fig:dispersion}a), and then locate the particles' center of mass (Fig. \ref{fig:dispersion}c). Next, we build the Delaunay network (Fig. \ref{fig:dispersion}d) to calculate $AD$. Fig. \ref{fig:areal disorder} shows the $AD$ for the composites in Fig. \ref{fig:microscopy images}. From Fig. \ref{fig:areal disorder}, we can see that the dispersion of our composites is random-like. 

\section*{S3: COMSOL simulations for obtaining eddy current damping forces}
\subsection*{S3.1. FEM: Eddy current damping in a graphite plate}
\begin{figure}[h]
    \centering
    \renewcommand{\thefigure}{S\arabic{figure}}
    \includegraphics[height=8cm]{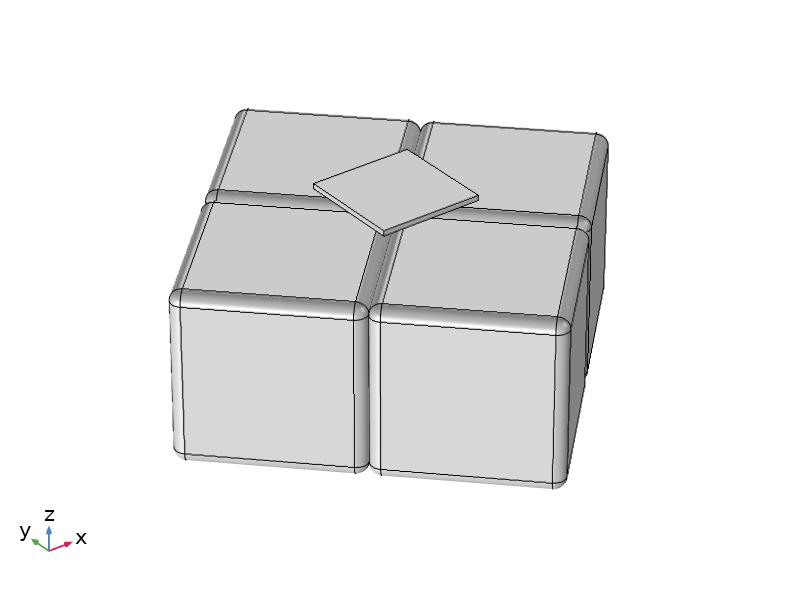}
    \caption{\label{fig:FEM geo model graphite}
    Geometry model of magnets and graphite plate.}
\end{figure}
This section details out the methodology we use to calculate the eddy current damping forces of a square graphite plate levitating above four permanent magnets. The geometry of the model is shown in Fig. \ref{fig:FEM geo model graphite}.

We simulate the magnetic field using \emph{COMSOL Multiphysics 5.6}. Assuming that the influence of the diamagnetic plate on the field is negligible, the integrated magnetic force on the plate, $\mathbf{F_B}$, can be determined using
\begin{equation}
    \label{FB}
    \mathbf{F_B}=\boldsymbol{\nabla} \int_\mathcal{V} \mathbf{M} \cdot \mathbf{B} \mathrm{d}\mathcal{V}=\frac{\mu_0}{2} \int_\mathcal{V} \boldsymbol{\nabla}(\chi_x H_x^2 + \chi_y H_y^2 +\chi_z H_z^2)\mathrm{d}\mathcal{V},
\end{equation}
where $\chi_x, \chi_y, \chi_z$ are the magnetic susceptibility of the levitating plate in $x, y, z$ directions, $\mathcal{V}$ is the volume of the plate, $\mathbf{B}$ is the magnetic field and $\mathbf{M}$ is the plate's magnetization. The components of the magnetic field inside the plate are $H_{x,y,z}=B_{x,y,z}/\mu$, where $\mu \approx \mu_0$ is the magnetic permeability of graphite. By calculating the magnetic force in $z$ direction with different levitation gaps between the plate and magnets, the levitation height where the $z$-component of magnetic force is equal and opposite to the gravitational force is obtained. Using the levitation height, the stiffness of the magnetic force $k$ can then obtained by taking the derivative of $F_\mathrm{z}$ over $z$ at the equilibrium point. Next, the resonance frequency $f_\mathrm{res}$ is calculated using $f_\mathrm{res}=\frac{1}{2\pi}\sqrt{\frac{k}{m}}$, where $m$ is the mass of the plate.

Then, we simulate the eddy current damping of the plate. We note that when a conductor moves with velocity vector $\mathbf{v}$ through a magnetic flux density field $\mathbf{B}$, the charge carriers inside the conductor feel an electric field $\mathbf{v}\times \mathbf{B}$ due to the Lorentz' force in addition to the field from the electric potential $V_e$, that generates an eddy current density $\mathbf{J}$ given by
\begin{equation}
	\mathbf{J}=-\mathbf{\sigma}\nabla V_e+\mathbf{\sigma}(\mathbf{v}\times \mathbf{B}),
	\label{eq: current density}
\end{equation}
where $\mathbf{\sigma}$ is the electrical conductivity. By combining Eq. (\ref{eq: current density}) with the current continuity condition $\nabla\cdot\mathbf{J}=0$ and the boundary condition $\mathbf{J}\cdot\mathbf{n}=0$ ($\mathbf{n}$ is the unit vector perpendicular to the boundary), we determine the eddy current density distribution $\mathbf{J}$  numerically for known $\mathbf{v}$, $\mathbf{\sigma}$, $\mathbf{B}$. We then evaluate the the total damping contribution due to eddy currents as follows
\begin{equation}
    \label{Eq: Fe}
    \mathbf{F_e}=\int_{\mathcal{V}} \mathbf{J}\times \mathbf{B} \mathrm{d}\mathcal{V},
\end{equation}
where integration is done over the graphite plate volume $\mathcal{V}$. Noting that the eddy current damping force $\mathbf{F}_{\rm e}$ is proportional and in the opposite direction of the velocity $v$, we then estimate the damping coefficient $c$. Finally, $Q$ of the plate can be obtained as
\begin{equation}
    \label{eq: Q}
    Q=\frac{2\pi m f_\mathrm{res}}{c}.
\end{equation}
All the parameters used in our simulations for pyrolytic graphite plates are listed in Table \ref{table: material properties of pg}.

\begin{table}
	\centering
	
	\caption{\label{table: material properties of pg} Material properties used for the simulations of the levitating pyrolytic graphite.}
	\begin{tabular}{|c|c|c|c|}
		\hline 
		Property & Symbol & Value & Unit \\ 
		\hline 
		Density & $\rho$ & 2070 & $\si{kg/m^3}$ \\ 
		\hline 
		Susceptibility $\perp$ \cite{simon2000diamagnetic}  & $\chi_z$ & -450 & $\times 10^{-6}$ \\ 
		\hline 
		Susceptibility $\parallel$ \cite{simon2000diamagnetic} & $\chi_{x,y}$ & -85 & $\times 10^{-6}$ \\ 
		\hline 
		Conductivity $\perp$\cite{pappis1961properties} & $\sigma_z$ & 200 & S/m \\ 
		\hline 
		Conductivity $\parallel$\cite{pappis1961properties} & $\sigma_{x,y}$ & 200000 & S/m \\ 
		\hline  
	\end{tabular} 
\end{table}

\subsection*{S3.2 Eddy current damping in a composite plate}
% \begin{figure}[!ht]

%     \centering
%     \includegraphics[height=6cm]{Figure_SI/SI_geo_model_composite.pdf}\\
%   \begin{textblock}{20}(-5,-2)
%      \huge sth
%     \end{textblock}

% \end{figure}

 \begin{figure}[h!]
    \centering
     \renewcommand{\thefigure}{S\arabic{figure}}
     \includegraphics[height=6cm]{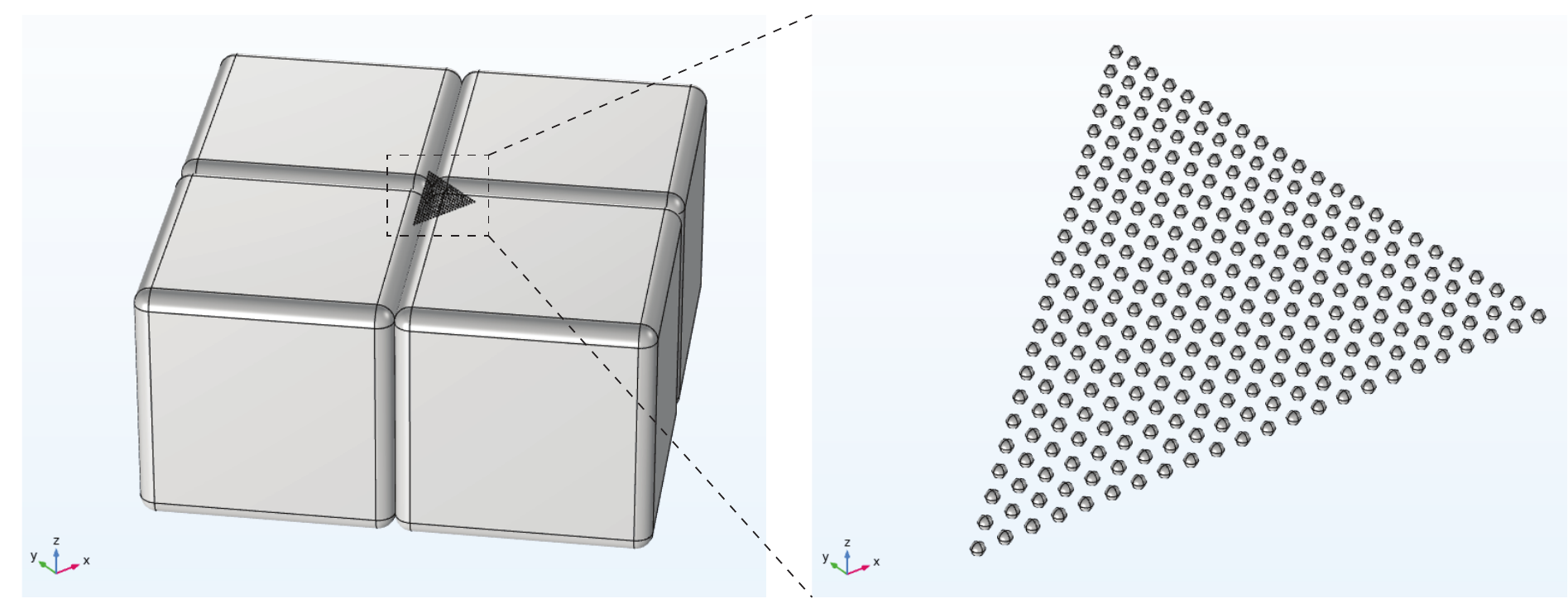}
    \caption{\label{fig:FEM geo model}
    COMSOL model. (a) Geometry model of magnets and graphite particles. (b) Spherical graphite particles.}
    \begin{textblock}{20}(-5.5,-3.8)
     (a) \hspace{7cm} (b)
    \end{textblock}
 \end{figure}
In this section we explain how the eddy current damping forces of a composite plate levitating above four permanent magnets are calculated. Because the magnetic susceptibility and electrical conductivity of the epoxy are negligible, the composite is modeled only by the graphite spheres to reduce the computation time, assuming that spheres are distributed homogeneously. As discussed in the main text, to account for experimental deviations from the theoretical model due to variations in particle size, composition, morphology and distribution, we use $C_\mathrm{r}d$ as the effective particle size in the simulation for particles with mean size of $d$, where $C_\mathrm{r}=6.3$ is an effective particle size factor. Fig. \ref{fig:FEM geo model}a shows the geometry model of four permanent magnets and 1/8 fraction of the graphite particles. 
The simulation procedure is similar to that of graphite plates.
The parameters used in our simulations for the composite plates are given in Table \ref{table: material properties of composite}.

\begin{figure}[h!]
    \centering
    \renewcommand{\thefigure}{S\arabic{figure}}
    \includegraphics[height=6cm]{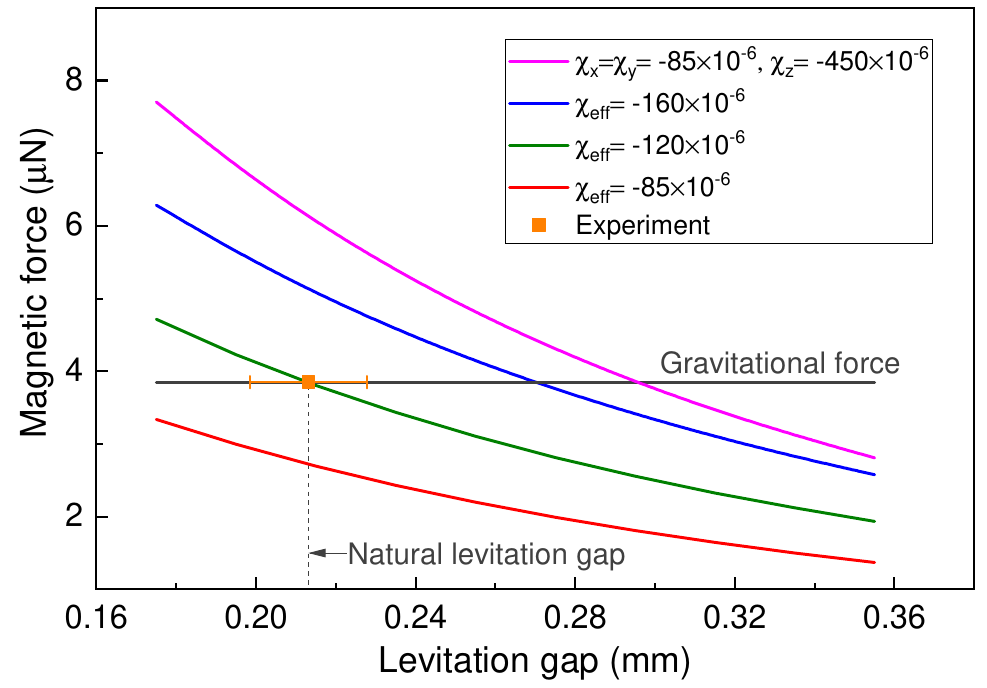}
    \caption{\label{fig:mag force}
    Magnetic force as a function of levitation gap obtained by COMSOL simulations using different susceptibilities (solid lines), and the measured natural levitation gap (dot).}
\end{figure}
We note that graphite is inherently anisotropic \cite{simon2000diamagnetic}. However, in our fabrication procedure graphite particles are randomly oriented in the epoxy matrix, and thus by considering all possible orientations in the matrix, the local anisotropy can be averaged out and the effective macroscopic behavior can be viewed isotropic. For this reason , in our study we treat the magnetic susceptibility of  graphite as an effective value $\chi_\mathrm{eff}$ which we evaluate by fitting our FEM simulations to the measured levitation height of the composite from experiments. In Fig. S8 we showcase how $\chi_\mathrm{eff}$ is estimated from experiments. We first measure the natural levitation gap of the composite using the Keyence microscope. By knowing the natural levitation gap and the gravitational force we then use our FEM model to estimate the magnetic force for different values of  $\chi_\mathrm{eff}$, and find the best value that fits the experimental finding. In Fig. S8 we show this procedure for a composite plate with $1.8\times1.8\times\SI{0.09}{mm^3}$, with volume fraction $V_\mathrm{f}=0.21$ and particle size $d=\SI{17.6}{\um}$. It can be seen that the green line which matches our experimental result is below the upper bound of the magnetic forces evaluated considering anisotropic magnetic susceptibilities, and the evaluated effective magnetic susceptibility $\chi_\mathrm{eff}=\num{-120e-6}$ agrees well with the reported value in reference \cite{boukallel2003levitated}. 

%\textcolor{red}{Fig. \ref{fig:mag force} shows the magnetic force as a function of levitation gap for a $1.8\times1.8\times\SI{0.09}{mm^3}$ composite plate with volume fraction $V_\mathrm{f}=0.21$ and particle size $d=\SI{17.6}{\um}$ obtained by COMSOL simulations when using different magnetic susceptibilities (solid lines).  The measured natural levitation gap is also shown in Fig. \ref{fig:mag force}, from which a good agreement between the simulation and measurement is observed when the effective susceptibility is $\chi=\SI{-120e-6}{}$.}

\begin{table}[h!]
	\centering
	\caption{\label{table: material properties of composite} Material properties of the graphite particles and epoxy used for the simulations of the levitating composite plates.}
	\begin{tabular}{|c|c|c|c|}
		\hline 
		Property & Symbol & Value & Unit \\ 
		\hline 
		Graphite density & $\rho_\mathrm{g}$ & 2250 & $\si{kg/m^3}$ \\
		\hline 
		Epoxy density & $\rho_\mathrm{e}$ & 1100 & $\si{kg/m^3}$ \\
		\hline 
		Graphite susceptibility & $\chi_\mathrm{eff}$ & -120 & $\times 10^{-6}$ \\ 
		\hline 
		Graphite resistivity\cite{pappis1961properties} & $\rho_\mathrm{r}$ & 1/200000 & $\Omega$ m \\ 
		\hline  
	\end{tabular} 
\end{table}

%\newpage
\begin{figure}[h!]
    \centering
    \renewcommand{\thefigure}{S\arabic{figure}}
    \includegraphics[height=3.8cm]{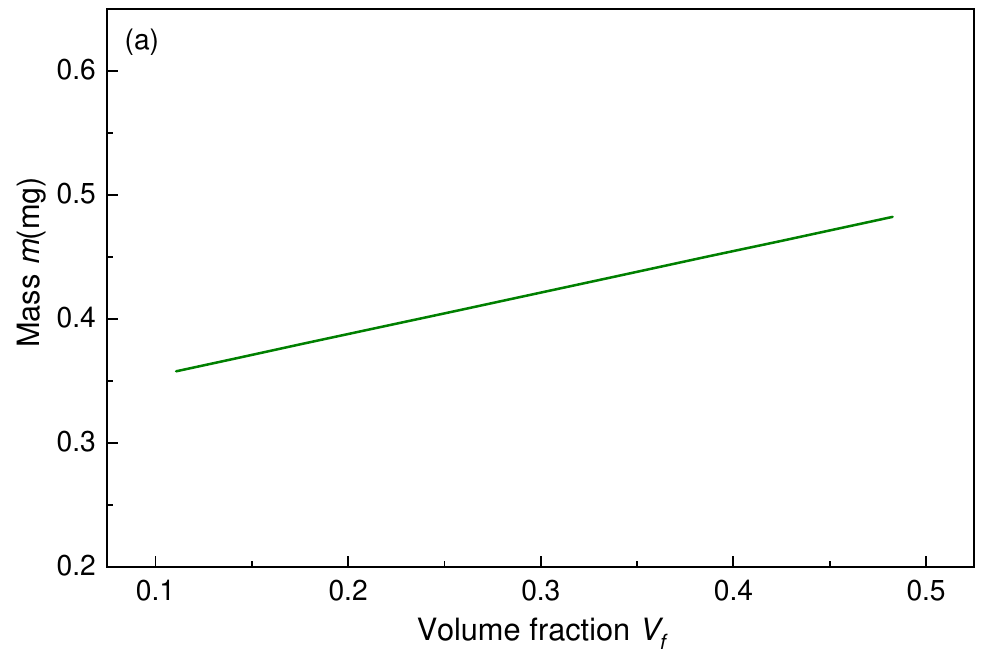}
    \includegraphics[height=3.8cm]{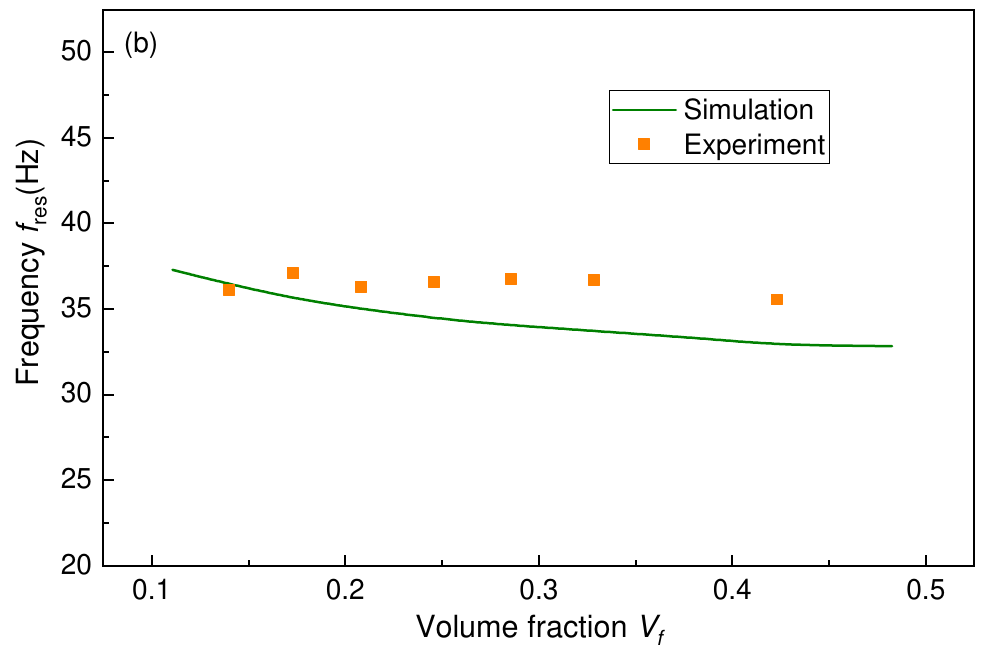}
    \includegraphics[height=3.8cm]{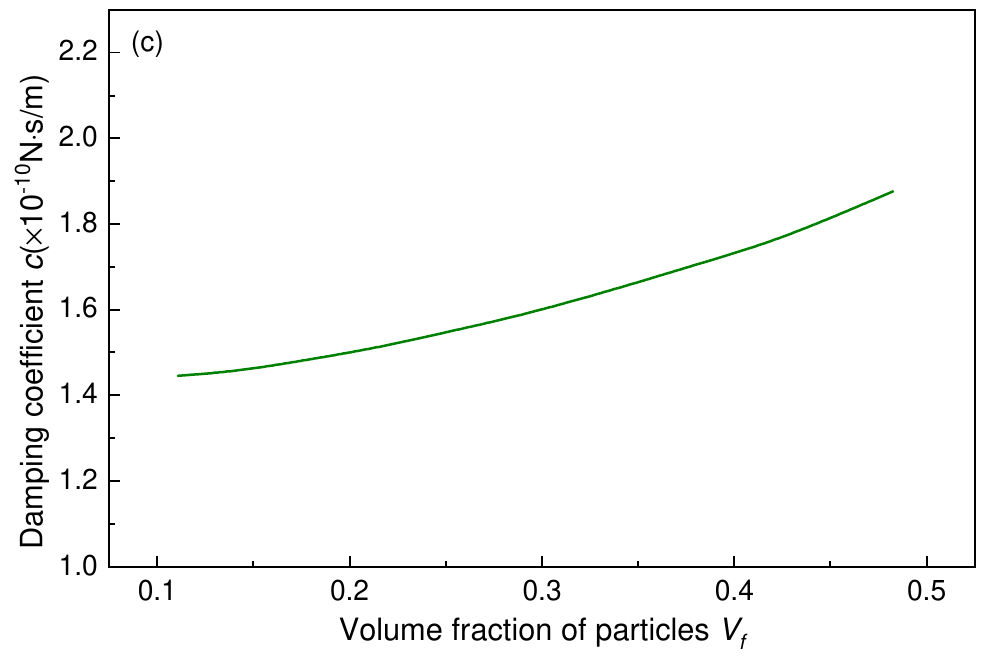}
    \includegraphics[height=3.8cm]{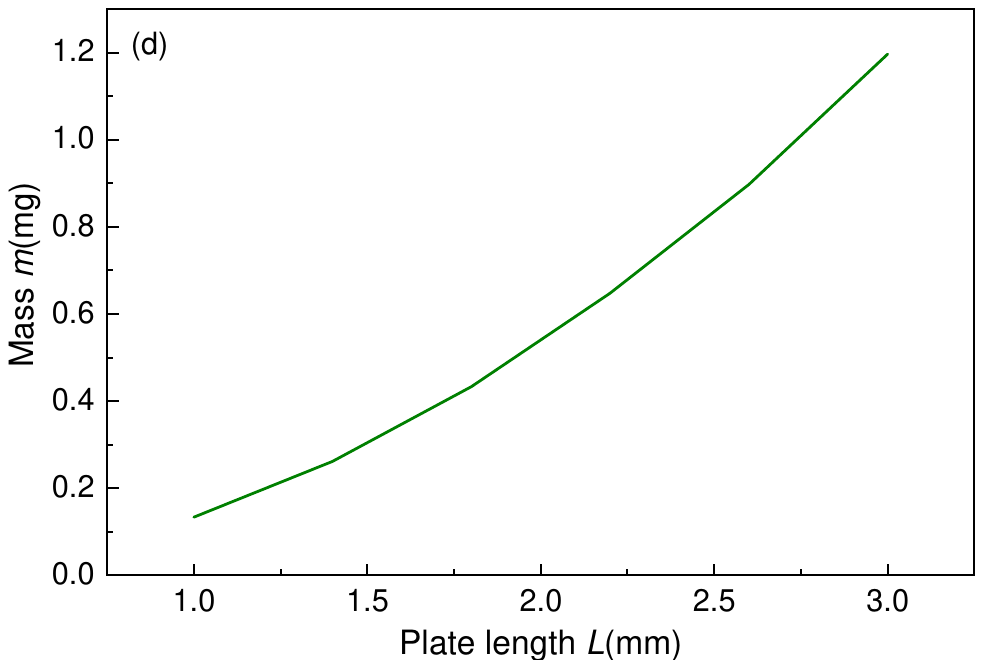}
    \includegraphics[height=3.8cm]{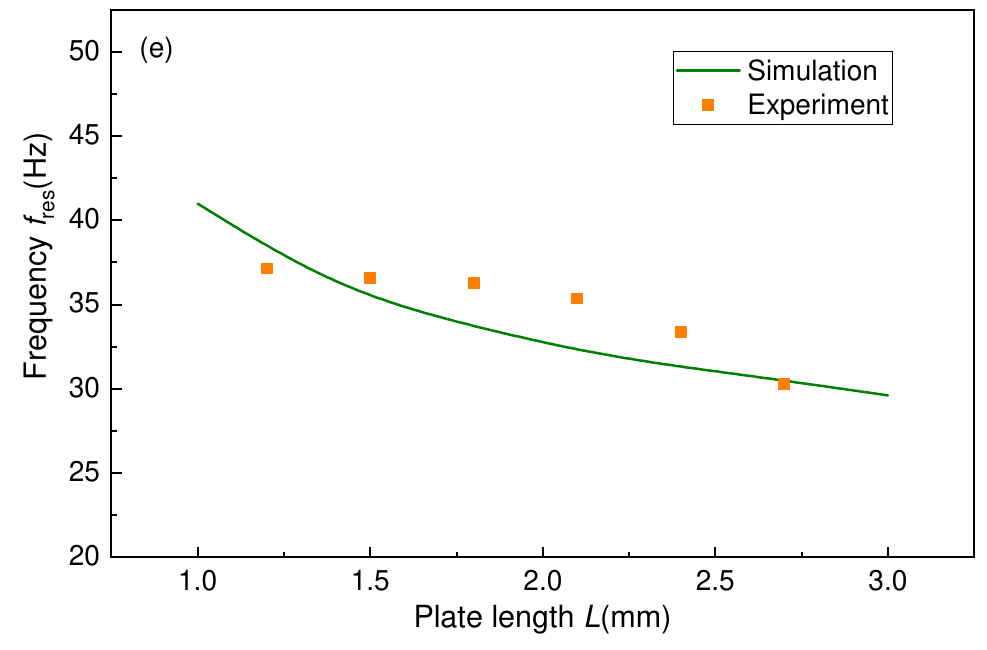}
    \includegraphics[height=3.8cm]{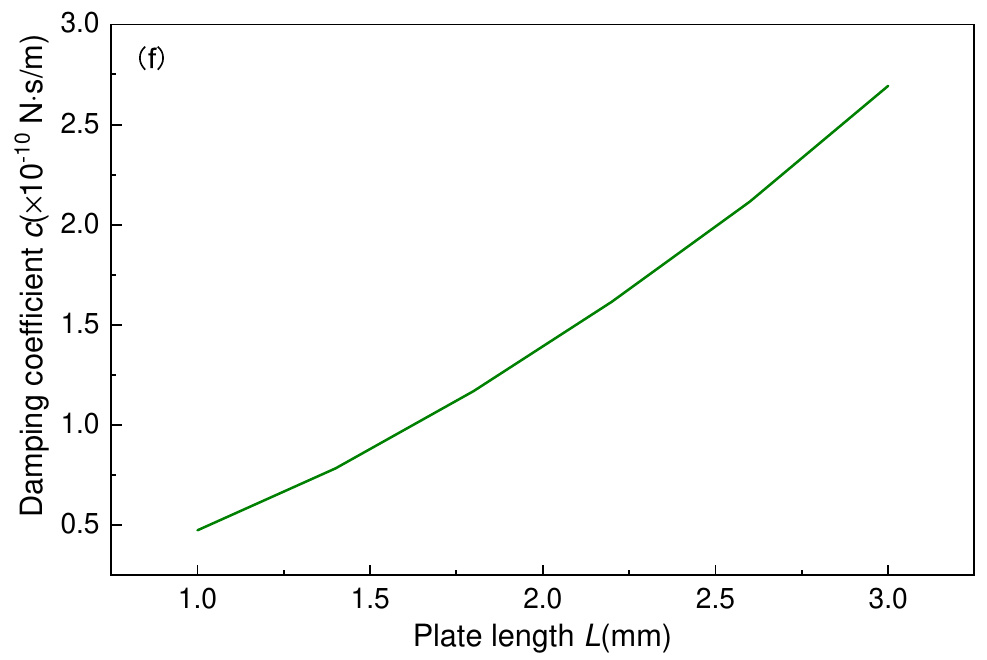}
    \includegraphics[height=3.8cm]{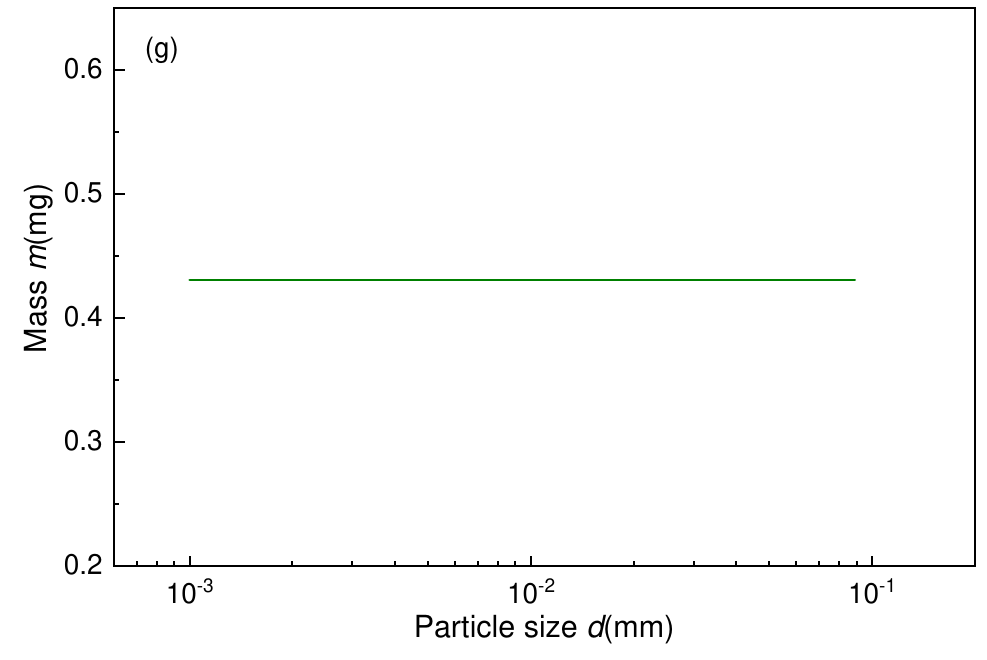}
    \includegraphics[height=3.8cm]{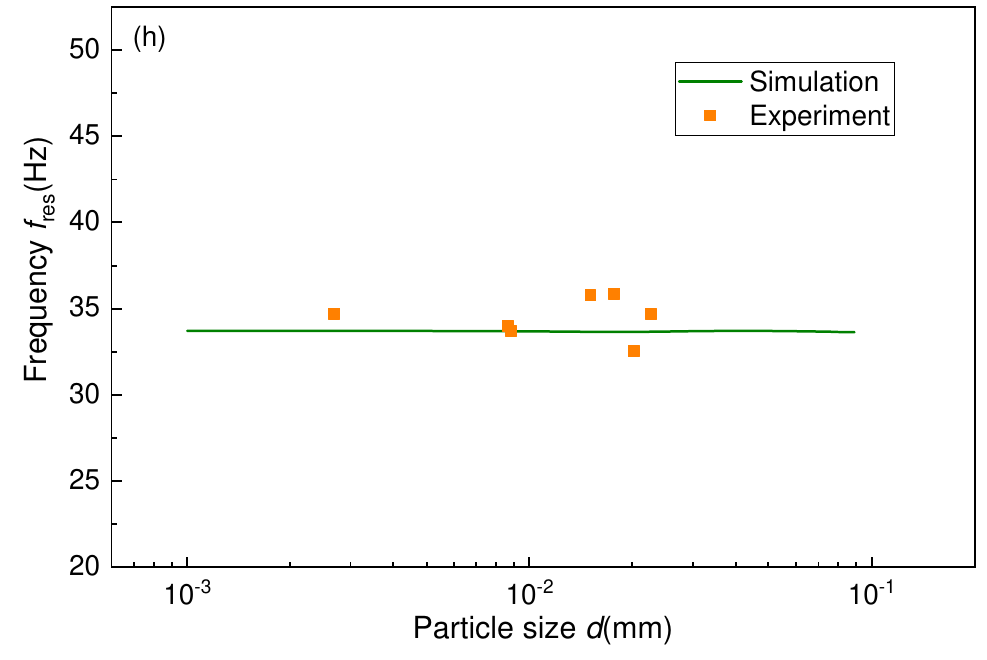}
    \includegraphics[height=3.8cm]{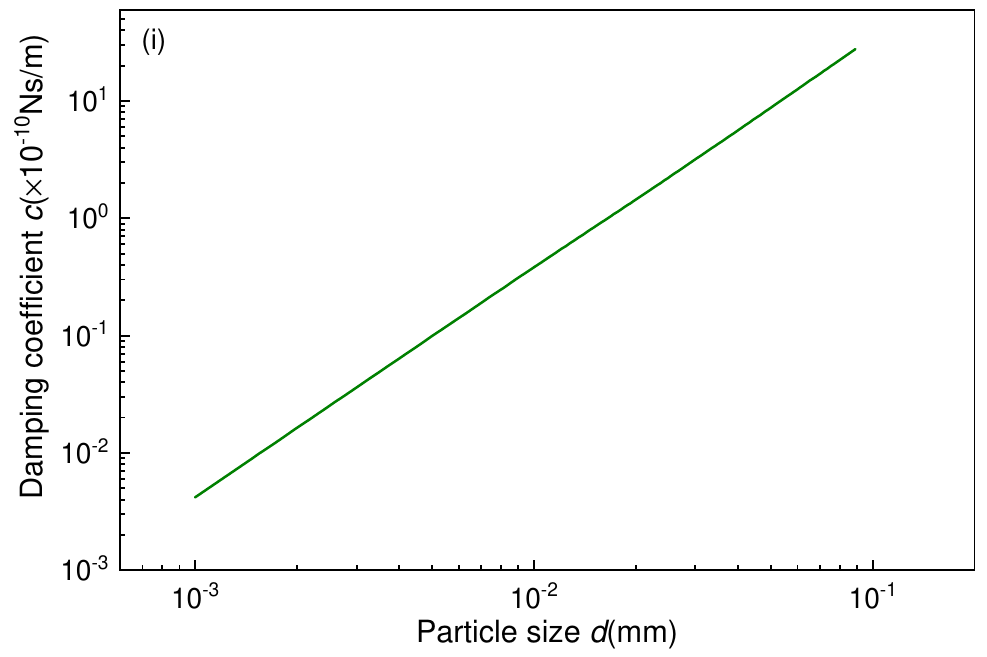}
    \caption{\label{fig: comsol result analysis}
    (a-c) Changes in mass $m$, resonance frequency $f_\mathrm{res}$ and damping coefficient $c$ of $1.8\times1.8\times\SI{0.09}{mm^3}$ composite plates with $d$=\SI{17.6}{\um} particles and different volume fraction $V_\mathrm{f}$; (d-f) Changes in mass $m$, resonance frequency $f_\mathrm{res}$ and damping coefficient $c$ of composite plates with $d$=\SI{17.6}{\um} particles and volume fraction $V_\mathrm{f}=0.21$, but different plate length $L$; (g-i) Changes in mass $m$, resonance frequency $f_\mathrm{res}$ and damping coefficient $c$ of $1.8\times1.8\times\SI{0.09}{mm^3}$ composite plates with volume fraction $V_\mathrm{f}=0.32$ and different particle size $d$. The lines correspond to data obtained from COMSOL simulations and dots represent measured data.}
    % \begin{textblock}{20}(-5,-7.5)
    %  (a) \hspace{5cm} (b) \hspace{4.8cm} (c)\\
    %  \vspace{3.8cm}
    %  (d) \hspace{4.8cm} (e) \hspace{4.8cm} (f)\\
    %  \vspace{3.7cm}
    %  (g) \hspace{4.8cm} (h) \hspace{4.8cm} (i)\\
    % \end{textblock}
\end{figure}
Fig. \ref{fig: comsol result analysis}a-c show the change of mass $m$, resonance frequency $f_\mathrm{res}$ and damping coefficient $c$ of composite plates with different particle volume fraction $V_\mathrm{f}$. Fig. \ref{fig: comsol result analysis}d-f show the change of mass $m$, resonance frequency $f_\mathrm{res}$ and damping coefficient $c$ of composite plates with different plate length $L$. Fig. \ref{fig: comsol result analysis}g-i show the change of mass $m$, resonance frequency $f_\mathrm{res}$ and damping coefficient $c$ of composite plates with different particle size $d$. 

\section*{S4: Electrical conductivity of the composites}
\begin{figure}[h!]
    \centering
    \renewcommand{\thefigure}{S\arabic{figure}}
    \includegraphics[height=6cm]{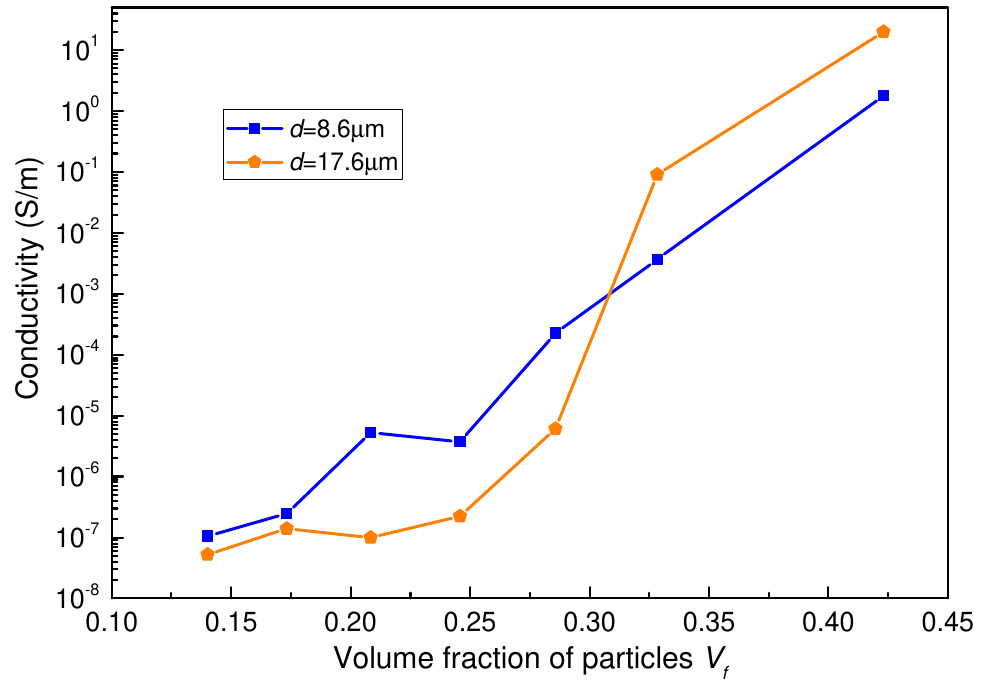}
    \caption{\label{fig:conductivity}
    Conductivity of two composites as a function of volume fraction.}
\end{figure}
Fig. \ref{fig:conductivity} shows the bulk conductivity of two composites made from $d=\SI{8.6}{\um}$ and $d=\SI{17.6}{\um}$ particles with different volume fractions. The conductivity is measured by the two-point measurement method with Agilent 4263B LCR meter (Santa Clara, CA, USA). It can be seen from the figure that the bulk conductivity of the composite is increasing with higher graphite volume fractions.

\section*{S5: Analytical modelling of eddy current damping}
%\subsection{$Q$ of a sphere}
\begin{figure}[h!]
    \centering
    \renewcommand{\thefigure}{S\arabic{figure}}
    \includegraphics[height=5cm]{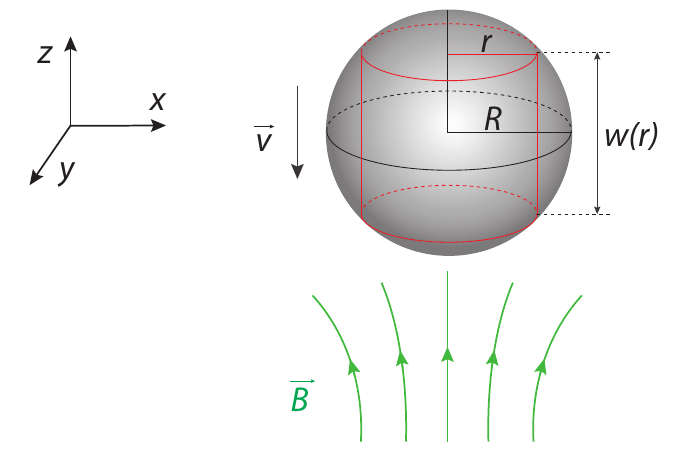}
    \caption{\label{fig:model}
    A sphere model.}
\end{figure}
In this section we obtain the $Q$-factor of a diamagnetic particle moving in a magnetic field, analytically.
We consider a spherical particle with radius $R=d/2$ that moves inside a magnetic field $\vv{B}$ as shown in Fig. \ref{fig:model}. To calculate the eddy current loss of the sphere, we assume the sphere consists of cylinders with varying radius $r$ and height $w(r)$ as shown in Fig. \ref{fig:model}. It is also assumed that the sphere is small compared to the magnetic field distribution and  that the magnetic field $\mathbf{B}$ is not changing in different locations on the sphere. According to Faraday's law of induction, the induced voltage on each ring (d$r$) of the sphere can then be calculated as \cite{taghvaei2009eddy}:
\begin{equation}
    \label{eq:emf}
    emf=\frac{\mathrm{d}\Phi}{\mathrm{d}t}=\frac{\mathrm{d}(\pi r^2B)}{\mathrm{d}t}=\pi r^2\frac{\mathrm{d}B}{\mathrm{d}z}\frac{\mathrm{d}z}{\mathrm{d}t}=\pi r^2v\frac{\mathrm{d}B}{\mathrm{d}z},
\end{equation}
where $\Phi$ is the magnetic flux, and $v$ is the velocity of the sphere. The induced current in the ring will then be
\begin{equation}
    \label{eq:dI}
    \mathrm{d}I=\pi r^2v\frac{\mathrm{d}B}{\mathrm{d}z}\frac{w(r)\mathrm{d}r}{2\pi r\rho_\mathrm{r}},
\end{equation}
where $\rho_\mathrm{r}$ is the electrical resistivity of the sphere. Using Eq. (\ref{eq:emf}) and (\ref{eq:dI}), the power loss in the ring can then be calculated as
\begin{equation}
    \mathrm{d}P=\pi r^3v^2\left(\frac{\mathrm{d}B}{\mathrm{d}z}\right)^2\frac{w(r)\mathrm{d}r}{2 \rho_\mathrm{r}},
\end{equation}
from which the total loss inside the sphere will become 
\begin{equation}
    P=\pi v^2\left(\frac{\mathrm{d}B}{\mathrm{d}z}\right)^2\frac{1}{2 \rho_\mathrm{r}}  \int_{0}^{R} r^32\sqrt{R^2-r^2} \mathrm{d}r=\frac{2\pi v^2R^5}{15\rho_\mathrm{r}} \left(\frac{\mathrm{d}B}{\mathrm{d}z}\right)^2.
\end{equation}
Finally, the eddy current loss per unit volume of the sphere becomes
\begin{equation}
    \label{eq: si unit power loss}
    P_\mathrm{unit}=\frac{v^2R^2}{10\rho_\mathrm{r}} \left(\frac{\mathrm{d}B}{\mathrm{d}z}\right)^2.
\end{equation}
Considering $P_\mathrm{unit}=F_\mathrm{eddy}v=c_\mathrm{unit}v^2$, and using Eq. \ref{eq: si unit power loss}, the damping coefficient $c_\mathrm{unit}$ can then be expressed as
%\begin{equation}
%    P_\mathrm{unit}=F_\mathrm{eddy}v=c_\mathrm{unit}v^2,
%\end{equation}
%combining with Eq. \ref{eq: si unit power loss}, the damping coefficient of unit volume $c_\mathrm{unit}$ can be expressed as
\begin{equation}
    c_\mathrm{unit}=\frac{R^2}{10\rho_\mathrm{r}} \left(\frac{\mathrm{d}B}{\mathrm{d}z}\right)^2.
\end{equation}
Which results in the following expression for the $Q$-factor solely due to eddy currents:
\begin{equation}
    \label{eq: SI-Q simple model}
    Q_\mathrm{sphere}=\frac{2\pi m_\mathrm{sphere}f_\mathrm{res}}{c_\mathrm{sphere}}=\frac{2\pi \rho_\mathrm{d}V_\mathrm{sphere}f_\mathrm{res}}{c_\mathrm{unit}V_\mathrm{sphere}}=\frac{2\pi \rho_\mathrm{d}f_\mathrm{res}}{c_\mathrm{unit}}=\frac{20\pi f_\mathrm{res}\rho_\mathrm{d}\rho_\mathrm{r}}{R^2(\mathrm{d}B/\mathrm{d}z)^2},
\end{equation}
in which $m_\mathrm{sphere}$ is the mass of the particle, $f_\mathrm{res}$ is the resonance frequency, $\rho_\mathrm{d}$ is the density and $V_\mathrm{sphere}$ is the volume of the sphere. 
%\subsection{Calculating eddy current damping in a composite plate}

For a composite plate consisting of spheres of radius $R$ dispersed in an insulating matrix, the $Q$ can be determined similar to Eq.(\ref{eq: SI-Q simple model}) as follows:
\begin{equation}
    \label{eq: SI-Q simple model 2}
    Q_\mathrm{plate}=\frac{2\pi m_\mathrm{plate}f_\mathrm{res}}{c_\mathrm{plate}}=\frac{2\pi \rho_\mathrm{p}V_\mathrm{plate}f_\mathrm{res}}{\int_{V_\mathrm{spheres}} c_\mathrm{unit}\mathrm{d}V_\mathrm{spheres}}=\frac{20\pi f_\mathrm{res}\rho_\mathrm{p}\rho_\mathrm{r}V_\mathrm{plate}}{R^2\int_{V_\mathrm{spheres}} \left(\frac{\mathrm{d}B}{\mathrm{d}z}\right)^2 \mathrm{d}V_\mathrm{spheres}},
\end{equation}
where $\rho_\mathrm{p}$ is the density of the composite plate, $V_\mathrm{plate}$ is the volume of the plate, and  $V_\mathrm{spheres}$ is the volume of the spherical particles. For a graphite/epoxy composite plate with volume fraction of $V_\mathrm{f}$, the $Q$ is then 
\begin{equation}
    \label{eq: SI-Q simple model 2}
    Q_\mathrm{plate}=\frac{20\pi f_\mathrm{res}\rho_\mathrm{r}V_\mathrm{plate}(V_\mathrm{f}(\rho_\mathrm{g}-\rho_\mathrm{e})+\rho_\mathrm{e})}{(C_\mathrm{r}R)^2\int_{V_\mathrm{spheres}} \left(\frac{\mathrm{d}B}{\mathrm{d}z}\right)^2 \mathrm{d}V_\mathrm{spheres}}=\frac{20\pi f_\mathrm{res}\rho_\mathrm{r}V_\mathrm{plate}(V_\mathrm{f}(\rho_\mathrm{g}-\rho_\mathrm{e})+\rho_\mathrm{e})}{(C_\mathrm{r}R)^2\nabla^2 B V_\mathrm{f}V_\mathrm{plate}}=\frac{20\pi f_\mathrm{res}\rho_\mathrm{r}((\rho_\mathrm{g}-\rho_\mathrm{e})+\rho_\mathrm{e}/V_\mathrm{f})}{(C_\mathrm{r}R)^2\nabla^2 B},
\end{equation}
where $\rho_\mathrm{g}$ is the density of graphite, $\rho_\mathrm{e}$ is the density of epoxy, and $C_\mathrm{r}$ is the apparent particle
size factor that accounts for the uncertainties related to
particle size, morphology and distribution. And
\begin{equation}
    \nabla^2 B=\frac{\int_{V_\mathrm{plate}}\left(\frac{\mathrm{d}B}{\mathrm{d}z}\right)^2\mathrm{d}V_\mathrm{plate}}{V_\mathrm{plate}},
\end{equation}
that can be obtained numerically using Comsol.

\section*{S6: $Q$-factors and acceleration noise floor of state-of-the-art levitodynamic systems}
In Table III - V we list the $Q$s and acceleration noise floors $\sqrt{S_\mathrm{aa}}$ of different levitodynamic systems showed in Fig. 5 of the main text. In Table III, the $Q$s were measured directly without feedback cooling either at room temperature or cryogenic temperature. In Table IV, we show the $Q$s measured at room temperature with feedback cooling and the natural $Q$s estimated at room temperature without feedback cooling. In Table V, the reported acceleration noise floor of different levitodynamic systems under different measurement conditions are listed. The data marked with $[*]$ are estimated using the following equation \cite{timberlake2019acceleration}:
\begin{equation}
    \sqrt{S_{aa}}=\sqrt{\frac{8\pi f_\mathrm{res} k_\mathrm{B} T}{mQ}}
    \label{Eq:sensitivity}
\end{equation}
where $f_\mathrm{res}$ is the resonance frequency, $m$ is the mass, $Q$ is the quality factor, $T$ is the temperature and $k_B$ is the Boltzmann constant.
\begin{table}[h!]
	\centering
	\caption{\label{table: Q-mass} $Q$-factors of different levitodynamic systems without feedback cooling (RT: room temperature).}
	\begin{tabular}{>{\centering\arraybackslash}m{2cm}cccc}
		\hline 
		mass(kg) & $Q$ & Levitation method & Temperature & Reference \\ 
		\hline
		\num{3.3e-18} & \num{1.0e7} & \multirow{2}{*}{optical} & RT & \cite{gieseler2012subkelvin}\\
		
		\num{3.7e-14} &  \num{2.1e4} & & RT & \cite{li2011millikelvin} \\
		\hline
		\num{9.6e-17} & \num{1.5e6} & electrical & RT & \cite{pontin2020ultranarrow} \\
		\hline
		\num{6.1e-10} & \num{1.3e7} & \multirow{4}{*}{superconducting} & 4.2K & \cite{vinante2020ultralow} \\
		%\hline
		\num{1.1e-10} &  \num{1.0e6} & & $<$90K & \cite{gieseler2020single} \\
		%\hline
		\num{5.7e-11} & \num{5.0e4} &  & 5K & \cite{wang2019dynamics} \\
		%\hline
		\num{4.0e-6} &  \num{5.5e3} & & 5K & \cite{timberlake2019acceleration} \\
		\hline
		\num{2.7e-13} & \num{2.0e7} & \multirow{7}{*}{diamagnetic} & 3K & \cite{leng2021mechanical} \\
		%\hline
		\num{7.8e-8} &  \num{1.5e5} & & RT & \cite{PhysRevApplied.16.L011003} \\
		%\hline
		\num{1.0e-5} &  \num{362} & & RT & \cite{chen2020rigid} \\
		%\hline
		\num{2.3e-5} &  \num{176} & & RT & \cite{chen2020rigid} \\
		%\hline
		\num{3.9e-5} &  \num{115} & & RT & \cite{chen2020rigid} \\
		%\hline
		\num{6.3e-5} &  \num{76} & & RT & \cite{chen2020rigid} \\
		%\hline
		\num{2.3e-6} &  \num{4.6e5} & & RT & this work \\
		\hline
	\end{tabular} 
\end{table}
%Q with feedback cooling 
\begin{table}[h!]
	\centering
	\caption{\label{table: Q-mass-fb} $Q$-factors of different levitodynamic systems with feedback cooling.}
	\begin{tabular}{>{\centering\arraybackslash}m{2cm}cc>{\centering\arraybackslash}m{0.1cm}cccc}
		\hline 
		\multirow{2}{*}{mass(kg)} &  \multicolumn{2}{c}{With feedback cooling} & &  \multicolumn{2}{c}{Estimated natural damping} & \multirow{2}{*}{Levitation method} & \multirow{2}{*}{Ref}\\\cline{2-3} \cline{5-6}
		& Q & Effective temperature && Q & Effective temperature &  & \\
		\hline
		\num{1.4e-19} & 440 & 3mK && \num{4.4e7} & 300K & optical & \cite{vovrosh2017parametric}\\
		%\hline
		\num{3.1e-17} & 38.6 & 460mK && \num{2.5e4} & 300K & optical & \cite{ranjit2016zeptonewton}\\
		%\hline
		\num{3.0e-14} & 14.8 & 10K && \num{445} & 300K & optical & \cite{ranjit2015attonewton}\\
		\hline
		\num{2.5e-10} & 175 & 9K && \num{5.8e3} & 300K & \multirow{2}{*}{diamagnetic} & \cite{lewandowski2021high}\\
		\num{3.1e-15} & 13.7 & 1.2mK && \num{3.2e6} & 295K &  & \cite{slezak2018cooling}\\		
		\hline
	\end{tabular} 
\end{table}
\begin{table}[h!]
	\centering
	\caption{\label{table: Saa-mass} \centering Acceleration noise floor of different levitodynamic systems showed in Fig. 5 of the main text. Data marked with $[*]$ represent the estimated value.}
	\begin{tabular}{cccccc}
		\hline 
		mass(kg) & $\sqrt{S_\mathrm{aa}}(\si{g/\sqrt{Hz}})$ & levitation method & Effective temperature & Reference & Note\\ 
		\hline 
		\num{3.3e-18} & \num{1.1e-3} & \multirow{11}{*}{optical} & RT & \cite{gieseler2012subkelvin} & $*$\\
		%\hline 
		\num{3.7e-14} & \num{1.2e-4} &  & RT & \cite{li2011millikelvin} & $*$\\
		%\hline 
		\num{1.4e-19} & \num{3.9e-3} &  & 3mK & \cite{vovrosh2017parametric} & feedback cooling, $*$\\
		%\hline 
		\num{3.1e-17} & \num{5.4e-3} &  & 460mK & \cite{ranjit2016zeptonewton} & feedback cooling\\
		%\hline
		\num{3.0e-14} & \num{7.0e-4} &  & 10K & \cite{ranjit2015attonewton} & feedback cooling\\
		%\hline
		\num{1.4e-18} & \num{2.3e-3} &  & 3K & \cite{hempston2017force} & feedback cooling\\
		%\hline 
		\num{8.4e-14} & \num{1.2e-5} &  & $\sim$ & \cite{kawasaki2020high} & feedback cooling\\
		%\hline
		\num{1.4e-13} & \num{3.6e-5} &  & 1K & \cite{moore2014search} & feedback cooling\\
		%\hline
		\num{1.5e-13} & \num{7.5e-6} &  & $\sim$mK & \cite{rider2018single} & feedback cooling\\
		%\hline
		\num{1.2e-11} & \num{4.0e-7} &  & $\sim$ & \cite{monteiro2017optical} & feedback cooling\\
		%\hline
		\num{9.4e-13} & \num{9.5e-8} &  & $\SI{50}{\micro K}$ & \cite{monteiro2020force} & feedback cooling\\
		\hline
		\num{9.6e-17} & \num{2.4e-5} & electrical & RT & \cite{pontin2020ultranarrow} & $*$\\
		\hline
		\num{6.1e-10} & \num{8.5e-10} & \multirow{4}{*}{superconducting} & 4.2K & \cite{vinante2020ultralow} & $*$\\
		%\hline
		\num{1.1e-10} & \num{5.0e-8} &  & $<$90K & \cite{gieseler2020single} & $*$\\
		%\hline
		\num{5.7e-11} & \num{1.6e-8} &  & 5K & \cite{wang2019dynamics} & $*$\\
		%\hline
		\num{4.0e-6} & \num{1.2e-10} &  & 5K & \cite{timberlake2019acceleration} & $*$\\
		\hline
		\num{2.7e-13} & \num{4.8e-9} & \multirow{9}{*}{diamagnetic} & 3K & \cite{leng2021mechanical} & $*$\\
		%\hline
		\num{7.8e-8} & \num{9.7e-10} &  & RT & \cite{PhysRevApplied.16.L011003} & -\\
		%\hline
		\num{2.5e-10} & \num{3.6e-8} &  & 9K & \cite{lewandowski2021high} & feedback cooling\\
		%\hline
		\num{3.1e-15} & \num{3.1e-6} &  & 1.2mK & \cite{slezak2018cooling} & feedback cooling, $*$\\
		%\hline
		\num{1.0e-5} & \num{2.6e-9} &  & RT & \cite{chen2020rigid} & $*$\\
		%\hline
		\num{2.3e-5} & \num{2.3e-9} &  & RT & \cite{chen2020rigid} & $*$\\
		%\hline
		\num{3.9e-5} & \num{2.1e-9} &  & RT & \cite{chen2020rigid} & $*$\\
		%\hline
		\num{6.3e-5} & \num{1.9e-9} &  & RT & \cite{chen2020rigid} & $*$\\
		%\hline
		\num{2.3e-6} & \num{1.6e-10} &  & RT & this work & $*$\\
		\hline
	\end{tabular} 
\end{table}

%\section{Summary}
\begin{comment}
\section*{S7: Magnetic susceptibility estimation of graphite particles}
\begin{figure}[h!]
    \centering
    \renewcommand{\thefigure}{S\arabic{figure}}
    \includegraphics[height=6cm]{Figure_SI/Magnetic-force-gap-different-chi.pdf}
    \caption{\label{fig:mag force}
    \textcolor{red}{Magnetic force as a function of levitation gap obtained by COMSOL simulations using different susceptibilities (solid lines), and the measured natural levitation gap (dot).}}
\end{figure}
\textcolor{red}{Fig. \ref{fig:mag force} shows the magnetic force as a function of levitation gap for a $1.8\times1.8\times\SI{0.09}{mm^3}$ composite plate with volume fraction $V_\mathrm{f}=0.21$ and particle size $d=\SI{17.6}{\um}$ obtained by COMSOL simulations when using different magnetic susceptibilities (solid lines).  The measured natural levitation gap is also shown in Fig. \ref{fig:mag force}, from which a good agreement between the simulation and measurement is observed when the effective susceptibility is $\chi=\SI{-120e-6}{}$.}

\end{comment}
%\section{Summary}

%\section*{Acknowledgments}

%\nocite{*}
\newpage
%\bibliography{ref-high-Q-si}% Produces the

\end{document}